\documentclass[USenglish,oneside,twocolumn]{article}

\usepackage[big]{dgruyter_NEW}
\usepackage[utf8]{inputenc}
\usepackage{todonotes}
\usepackage{hyperref}
\usepackage{multirow}
\usepackage{booktabs,enumitem}
\usepackage{float,listings,graphicx,color}
\usepackage{amsfonts,amsmath}
\usepackage{tikz}
\usepackage{tikzpeople}
\usetikzlibrary{patterns,arrows}
\usepackage{pgfplots,pgfplotstable}
\pgfplotsset{width=7cm,compat=1.15}

\def\isfullversion{1}

\usepackage{ifthen}
\newcommand{\fullversion}[2]{
  \ifthenelse{\equal{\isfullversion}{1}}
  {{#1}}
  {{#2}}
}

\usepackage{subcaption}
\usepackage{microtype}
\usepackage{natbib}
\usepackage{bm}

\usetikzlibrary{arrows,matrix,positioning,fit}

\usepackage{todonotes}

\newcommand\sign{\ensuremath{\mathrm{Sign}}}

\newcommand{\quant}{\mathsf{quant}}
\newcommand{\dequant}{\mathsf{dequant}}

\newcommand{\relu}{\ensuremath{\mathsf{ReLU}}}

\newcommand{\clamp}{\ensuremath{\mathsf{clamp}}}

\newcommand{\trunc}{\ensuremath{\mathsf{Trunc}}}
\newcommand{\truncp}{\ensuremath{\mathsf{TruncPriv}}}
\newcommand{\truncpr}{\ensuremath{\mathsf{TruncPr}}}

\newcommand\R{\ensuremath{\mathbb{R}}}
\newcommand\F{\ensuremath{\mathbb{F}}}
\newcommand\Z{\ensuremath{\mathbb{Z}}}
\newcommand\N{\ensuremath{\mathbb{N}}}
\newcommand\Ztk{\ensuremath{\mathbb{Z}_{2^k}}}

\newcommand\set[1]{\ensuremath{\lbrace #1\rbrace}}
\renewcommand\b[1]{\ensuremath{\lowercase{\bm{#1}}}}

\newcommand{\round}[1]{\left\lfloor#1 \right\rceil}
\newcommand{\floor}[1]{\left\lfloor#1 \right\rfloor}
\newcommand{\shr}[1]{\left\langle #1\right\rangle}

\tikzstyle{block} = [rectangle, minimum width=2cm, minimum height=1cm, text centered, draw=black, fill=red!15]
\tikzstyle{blockA} = [rectangle, minimum width=2cm, minimum height=1cm, text centered, draw=black, fill=blue!15]
\tikzstyle{blockB} = [rectangle, minimum width=2cm, minimum height=1cm, text centered, draw=black, fill=green!15]
\tikzstyle{arrow} = [thick,->,>=stealth]

\usepackage{xspace}
\newcommand{\mascot}{\ensuremath{\mathsf{MASCOT}}}
\newcommand{\lowgear}{\ensuremath{\mathsf{LowGear}}}
\newcommand{\spdztk}{\ensuremath{\mathsf{SPDZ2k}}}
\newcommand{\otsemitk}{\ensuremath{\mathsf{OTSemi2k}}}
\newcommand{\otsemip}{\ensuremath{\mathsf{OTSemiPrime}}}
\newcommand{\repltk}{\ensuremath{\mathsf{Replicated2k}}}
\newcommand{\replp}{\ensuremath{\mathsf{ReplicatedPrime}}}

\newcommand{\psrepltk}{\ensuremath{\mathsf{PsReplicated2k}}}
\newcommand{\psreplp}{\ensuremath{\mathsf{PsReplicatedPrime}}}

\usepackage{tcolorbox}
\tcbuselibrary{skins,breakable}

\newtcolorbox{protocol}[2][]{%
  enhanced,
  title        = {#2},
  attach boxed title to top left={xshift=+3mm,yshift*=-3mm},
  breakable    = true,
  colback      = black!4,
  colframe     = black!75,
  fonttitle    = \bfseries,
  colbacktitle = black!10!white,
  coltitle     = black,
  #1
}

\newtcolorbox{mynote}[1][]{%
  enhanced jigsaw, 
  breakable        = true,
  borderline west  ={2pt}{0pt}{red}, 
  sharp corners, 
  boxrule          =0pt, 
  fonttitle        ={\large\bfseries},
  coltitle         ={black},  
  colback          =yellow!25,
  title            ={Review:\ },  
  attach title to upper, 
  #1
}

\newtcbox{\xmybox}[1][red]{on line,
arc=7pt,colback=#1!10!white,colframe=#1!50!black,
before upper={\rule[-3pt]{0pt}{10pt}},boxrule=1pt,
boxsep=0pt,left=6pt,right=6pt,top=2pt,bottom=2pt}

\title{\huge Secure Evaluation of Quantized Neural Networks}
\runningtitle{Secure Evaluation of Quantized Neural Networks}

\cclogo{\includegraphics{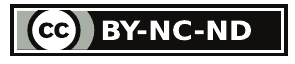}}

\begin{document}

  \author*[1]{Anders Dalskov}
  \author*[2]{Daniel Escudero}
  \author*[3]{Marcel Keller}
  \affil[1]{Aarhus University, E-mail: anderspkd@cs.au.dk}
  \affil[2]{Aarhus University, E-mail: escudero@cs.au.dk}
  \affil[3]{CSIRO's Data61, E-mail: mks.keller@gmail.com}

\begin{abstract}
  {We investigate two questions in this paper:
    \emph{First}, we ask to what extent ``MPC friendly'' models are already supported by major Machine Learning frameworks such as TensorFlow or PyTorch.
    Prior works provide protocols that only work on fixed-point integers and specialized activation functions, two aspects that are not supported by popular Machine Learning frameworks,
    and the need for these specialized model representations means that it is hard, and often impossible, to use e.g., TensorFlow to design, train and test models that later have to be evaluated securely.
    \emph{Second}, we ask to what extent the functionality for evaluating Neural Networks already exists in general-purpose MPC frameworks.
    These frameworks have received more scrutiny, are better documented and supported on more platforms.
    Furthermore, they are typically flexible in terms of the threat model they support.
    In contrast, most secure evaluation protocols in the literature are targeted to a specific threat model and their implementations are only a ``proof-of-concept'', making it very hard for their adoption in practice.
    We answer both of the above questions in a positive way:
    We observe that the quantization techniques supported by both TensorFlow, PyTorch and MXNet can provide models in a representation that can be evaluated securely; and moreover, that this evaluation can be performed by a general purpose MPC framework.
    We perform extensive benchmarks to understand the exact trade-offs between different corruption models, network sizes and efficiency.
    These experiments provide an interesting insight into cost between active and passive security, as well as honest and dishonest majority.
    Our work shows then that the separating line between existing ML frameworks and existing MPC protocols may be narrower than implicitly suggested by previous works.
    \\ \emph{\textbf{Summary of Changes}: Since the first version,
      figures for full implementation of inference of various networks
      in MPC have been added . Furthermore, Assi Barak
      has been removed from the authors at his request.}
  }

\end{abstract}

\maketitle
\paragraph*{Acknowledgements}

This work has been supported by the European Research Council (ERC) under the European Unions’s Horizon 2020 research and innovation programme under grant agreement No 669255 (MPCPRO),
and the Danish Independent Research Council under Grant-ID DFF-6108-00169 (FoCC) and
the European Union's Horizon 2020 research and innovation programme under grant agreement No 731583 (SODA).

We thank Adri\`a Gasc\'on for fruitful discussions, as well as members of NEC and BIU, especially Prof. Benny Pinkas and Prof. Yehuda Lindell.
We thank Assi Barak (who was previously on this paper) for many helpful discussions in the initial phases of this project.

\section{Introduction}
\label{sec:introduction}

Machine Learning (ML) models are becoming more relevant in our day-to-day lives due to their ability to perform predictions on several types of data.
Neural Networks (NNs), and in particular Convolutional Neural Networks (CNNs), have emerged as a promising solution for many real-life problems such as facial recognition \cite{facerecogCNN}, image and video analysis for self-driving cars \cite{DBLP:journals/corr/BojarskiTDFFGJM16} and even for playing games (most readers probably know of \emph{AlphaGo}~\cite{silver2016mastering} which in 2016 beat one of the top Go players). CNNs have also found applications within areas of medicine. \cite{esteva2017dermatologist}, for example, demonstrates that CNNs are as effective as experts at detecting skin cancers from images, and \cite{dunnmon2018assessment} investigated using CNNs to examine chest x-rays.

Many applications that use Machine Learning to infer something about a piece of data, does so on data of sensitive nature, such as in the two examples cited above.
In such cases the ideal would be to allow the input data to remain private.
Conversely, and since model training is by far the most expensive part of deploying a model in practice,%
\footnote{For example, the network by Yang et al. \cite{DBLP:journals/corr/abs-1906-08237} costs between \$$61\,000$ and \$$250\,000$ to train according to \url{https://syncedreview.com/2019/06/27/the-staggering-cost-of-training-sota-ai-models/}.} %
preserving model privacy may be desirable as well.

In order to break this apparent contradiction (performing computation on data that is ought to be kept secret) tools like \emph{secure multiparty computation} (MPC) can be used.
Using such tools, image classification can be performed so that it discloses neither the image to the model owner, nor the model to the input owner.
In the client-server model this is achieved by letting the data owner and the model owner \textit{secret-share} their input towards a set of servers, who then run the computation over these shares.

\subsection{Towards Deploying Secure Inference.}

Research in the area of secure evaluation of CNNs has been rich during the last couple of years \cite{DBLP:conf:icml:Gilad-BachrachD16,ASIACCS:RWTSSK18,DBLP:conf:dac:RouhaniRK18,CCS:LJLA17,SP:MohZha17,DBLP:conf/uss/JuvekarVC18,DBLP:journals/popets/WaghGC19,riazi2019xonn,cryptoeprint:2019:1049}.
The main goal of this prior research has been to reduce the performance gap between evaluating a CNN in the clear and doing it securely.
Current state of the art solutions rely on for example Garbled Circuits \cite{riazi2019xonn} or MPC \cite{cryptoeprint:2019:1049}.
Both of these works manage to evaluate large ImageNet type models (tens of layers and 1000 classes) with reasonable efficiency.
Moreover, the CrypTFlow framework \cite{cryptoeprint:2019:1049} also support secure inference with malicious security albeit by relying on a secure hardware assumption.
Several solutions have also been developed by researchers closer to the industry side, such as TF-Encrypted \cite{tfencrypted} or CrypTen \cite{crypten}, which suggests that secure prediction has value beyond the academic point of view---an important factor for accelerating wide adoption of these techniques.

These advances mean that secure evaluation of large models (tens of layers, millions of parameters) can now be performed in the order of seconds which, while too slow for real-time classification, is acceptable for many privacy critical tasks like the ones described in \cite{dunnmon2018assessment,esteva2017dermatologist}.
Medical tests typically take hours if not days or weeks, so running an additional test that takes a few seconds or minutes will not matter.

Given the intense interest in secure inference in recent years, one might ask: \emph{what obstacles are preventing the use of secure inference in practice?}.
Most of the focus on the research literature has been on improving performance, but this is far from being the only challenge in this direction.

\paragraph*{Challenges, the ML perspective.}

Setting aside the privacy requirement for a minute, deploying a Machine Learning solution in practice is already a long and strenuous process.
Data has to be acquired and then processed; a model has to be designed, its parameters have to be tuned and finally the model needs to be trained.
Moreover, the process is often repeated totally or in parts whenever new data is acquired or a better model design is found.

It is not surprising that a \emph{significant} amount of effort is being spent on designing feature-rich and well documented frameworks for developing these models, for training them and for testing them.
Frameworks such as TensorFlow \cite{tf}, PyTorch \cite{pt} and MXNet \cite{mx} are all seeing significant use and are being actively developed by major companies (Google for TensorFlow, Facebook for PyTorch and Apache for MXNet).

Ideally, it should be possible to design a model using these widely used frameworks and then simply alter the way it is used at the very end when the model is used to perform predictions on user provided inputs.
However, most existing solutions for secure inference throws a wrench into this process as they rely either on models that use specialized activation functions, such as CryptoNets \cite{DBLP:conf:icml:Gilad-BachrachD16} or require a specialized training process, such as XONN \cite{riazi2019xonn}.
For these reasons, a part that is often of the \emph{least} concern when developing a Machine Learning solution (using the final model for prediction) thus becomes an aspect that has to be considered at virtually all steps of the design process.

It is worth noting here that it is not enough for a framework to simply support conversion from a trained model into a representation that can be used.
Such conversions typically involve steps like moving from a floating point representation to a fixed point one, or exchanging certain activation functions with ``approximations''---all of which impact accuracy and thus the expressiveness of the model.

These issues were also identified by the authors of CrypTFlow \cite{cryptoeprint:2019:1049} and a third of their work is spent detailing a customized application that converts TensorFlow code into a representation which can be run securely by their framework, without substantially compromising on accuracy.

Our work takes a different approach to this problem.
Namely, we investigate whether, and to what extent, existing frameworks such as TensorFlow support training and designing models that can be securely evaluated.

\paragraph*{Challenges, the MPC perspective.}
In spite of notable progress in general-purpose MPC, most of the existing MPC-based secure inference works rely on customized subprotocols that are highly optimized for particular activation functions.
Moreover, and as remarked on above, these activation functions are often themselves novel in the sense that they rarely see use outside of secure inference.

Such a tight coupling between what can be evaluated and how its evaluated also implies that often only a single threat model is supported.
In particular, if a specific threat model is desired, then often this directly determines what kind of network that can be run.
For example, if one requires a dishonest majority solution, such as a 2-party protocol, then XONN is the current most efficient solution.
But XONN only works on binarized networks and only works with $\sign$ as the activation function.

At the other end, if one is fine with honest majority (typically 3 parties, as that is the most efficient setting) but require active security, then only CrypTFlow fits the bill; but CrypTFlow relies on secure hardware for active security.

Ideally, the MPC that is used should be ``oblivious'' to the network being evaluated, as that would allow a user to more freely choose which threat model is best suited for them \emph{without} having to think about the structure of the network, or specialized hardware.
This is the approach we take.
That is, we investigate how applicable general purpose MPC frameworks such as MP-SPDZ~\cite{cryptoeprint:2020:521} or SCALE-MAMBA \cite{scalemamba} are to evaluate Convolutional Neural Networks.

Using general-purpose MPC frameworks is not only convenient when it comes to the threat model.
Another important factor is that, given their flexibility, this type of protocols tend to receive more scrutiny from part of the community \cite{hastings2019sok}, they are much better understood from a practical point of view and they have more reference implementations.\footnote{Although MPC protocols, general-purpose or not, are accompanied with security proofs, this does not mean that their ``level of security'' is the same. Details that appear only at implementation time, like instantiations of random oracles, make it important to have an end-to-end understanding of the security. Having reference implementations that are as close to industry-grade as possible is important, and the panorama in this regard for general-purpose protocols is much more promising than for special-purpose MPC.}
These considerations are important for an area like MPC, which is today in a stage in which many applications are within the practical realm, but no standardization of implementation practices (like the ones found e.g.~in symmetric-key cryptography) are set yet.

\subsection{Our Contribution}
\label{sec:our-contribution}

Our work addresses the challenges identified in the previous section, and to this end it focuses on the following two questions that, while simple in nature, have so far not been treated in research on secure inference.

\begin{enumerate}
\item
  To what extent can ``MPC friendly'' models be obtained from existing frameworks such as TensorFlow or PyTorch, \emph{without} requiring a customized conversion protocol?
  More precisely, is it possible to design a model using these standard frameworks, which can then be \emph{efficiently} evaluated by a secure protocol, without at all tampering with the model?

\item
  To what extent does existing MPC frameworks support running models ``out-of-the-box''?
  That is, can we securely evaluate Machine Learning models (of the kind described above) using general-purpose MPC frameworks?
\end{enumerate}

These questions provide a minimal baseline that one should keep in mind before moving on to specialized protocols or resorting to specialized models that improve efficiency.
In this work we explore these questions thoroughly, which results in the following contributions:

\begin{description}
\item[Quantization.]
  We identify the quantization techniques used in both TensorFlow, PyTorch and MXNet and described in \cite{DBLP:journals/corr/abs-1712-05877} as particularly well suited for MPC.
  This type of quantization results in models for which each output of each convolutional layer can be expressed as a single dot-product followed by a truncation.
\item[MPC.]
  We describe how to implement these types models in a black-box way; that is, without resorting to special properties of the underlying MPC.
\item[Optimizations.]
  In settings where dot-products can be securely computed with high efficiency, the main bottleneck is truncation (or bitwise right-shift).
  As optimizations, we therefore present an optimized truncation protocol for some threat models which further improves efficiency.
\item[Experiments.]
  Finally, we evaluate the efficiency of a large class of quantized models in a variety of different threat models.
  More precisely, we evaluate 16 different models of varying size, each in 16 different settings.
\end{description}

We elaborate on each contribution below.

First, the fact that we identify a widely used quantization scheme as being particularly well suited for secure inference, provides a very promising area of study for both researchers and practitioners.
For researchers, it provides a fixed target for secure protocol design.
For practitioners, it shows that one does not have to abandon widely used Machine Learning frameworks in order to design models for practical use that can evaluated securely.

Secondly, showing that these techniques are compatible with the arithmetic black-box model (that is, only secure additions and multiplications are required) allows us to lower the requirements that an MPC protocol should satisfy in order to be suitable for secure inference.
This in effect allows us to extend the amount of protocols supported.
However, this does not mean that these protocols cannot be optimized, which in fact takes us to our third contribution: We present optimized primitives for the case of truncation over the ring $\Z_{2^k}$, which is not as well studied as the correponding problem over fields and constitutes the main bottleneck when securely evaluating our quantized neural networks.

Finally, to illustrate the advantages of our approach with respect to constructing ad-hoc MPC protocols that are only suitable to certain type of models, we perform a large amount of experiments with a wider range of models and different MPC protocols.
More precisely, we securely evaluate 16 models that are part of the ImageNet family\footnote{The choice of evaluating these models is only because of convenience, as they can be found as part of the Tensorflow Lite model repository. As discussed in Section \ref{sec:impl-benchm}, \emph{any} model trained with Tensorflow can be evaluated using our framework without the need of introducing extra tools.} using MPC protocols that vary with respect to several dimensions: Corruption threshold (honest vs.~dishonest majority), corruption model (passive vs.~active security) algebraic structure (integers modulo $2^{k}$ or modulo prime $p$) and whether or not truncation is exact or probabilistic.

Our experiments let us conclude several things:
\begin{enumerate}
\item Corruption threshold has a very large impact on efficiency for general-purpose MPC.
  Indeed, a dishonest majority evaluation takes orders of magnitude longer than honest majority.
\item On the other hand, corruption model has a comparatively smaller impact on efficiency. For example, an actively secure evaluation with an honest majority protocol over a prime field is only about 4 times slower than if the evaluation was done with the corresponding passive protocol.
\item For passive protocols we find that modulo a power of 2 is between 4 and 10 as efficient as the corresponding protocol modulo a prime power.
  This result further supports all the recent work that has gone into designing fast protocols that work over a ring \cite{defksv,C:CDESX18} (as this ring is typically taken to be integers modulo a power of 2).
\item Finally, by running our experiments for both exact truncation (meaning we evaluate the model \emph{exactly} as would be done in the clear) and probabilistic truncation (meaning the evaluation may suffer some unknown loss in accuracy) we can quantify the exact gain in efficiency by relying on specialized protocols.
\end{enumerate}

\subsection{Related Work}
\label{sec:related-work}
The following review is focused the different types of quantization (if any) that prior works has used; additionally, we look at frameworks for secure inference that prior works have developed.
A broader review can be found in Appendix \ref{sec:related-work-secure}.

\subsubsection{Quantization in prior work}
Whether implicitly or explicitly, most prior work already uses some form of quantization.
For instance, replacing floating-point by fixed-point numbers can be seen as a form of quantization.
More often than not, however, this conversion is done in a very naive manner where the primary goal has been to fit the model parameters to the secure framework without further consideration about any potential impact it might have on the model's accuracy.
Relatively little work has made explicit use of quantization in the context of securely evaluating Machine Learning models.
One example is the recent work by Bourse et al.~\cite{C:BMMP18}, where the authors use a quantization technique that is similar to the one described by Courbariaux and Bengio \cite{DBLP:journals:corr:CourbariauxB16}. Sanyal et al.~\cite{sanyal2018tapas} use the same techniques.
Nevertheless, their work lies in the FHE domain, which differs from multiparty computation.
For instance, the fact that the weights are kept in the clear by the model owner changes the way the computation is performed, and allows them to use only additions and subtractions.
XONN \cite{riazi2019xonn}, which is based on Garbled Circuits, uses a quantization scheme which converts weights into bits \cite{hubara2016binarized}.
For this to work, the authors need to increase the number of neurons of the network and a large part of their work is dedicated to describing how this scaling can be performed.
CrypTFlow \cite{cryptoeprint:2019:1049} employ what can be seen as a custom fixed-point-to-floating-point conversion protocol (called Athos) that automatically converts the floating point weights of a Tensorflow model into a fixed points representation, where the parameters are chosen so as to not compromise on the models original accuracy.

\subsubsection{Frameworks for secure evaluation}
Several previous works provide what can be viewed as a more complete framework for secure evaluation.
The first of these is MiniONN \cite{CCS:LJLA17} which provides techniques for converting existing models into models that can be evaluated securely.
The authors demonstrate this framework by converting and running several models for interesting problem domains, such as Language modeling, as well as more standard problems such as hand writing recognition (MNIST) and image recognition (CIFAR10).
CrypTFlow \cite{cryptoeprint:2019:1049} also provides more complete framework.
As already mentioned above, the first step in their framework is a protocol for converting an Tensorflow trained model into a model that can later be evaluated securely using a protocol based on SecureNN \cite{DBLP:journals/popets/WaghGC19}.

\subsection{Outline of the Document}
\label{sec:outline-document}

In Section \ref{sec:quantization} we give a brief introduction to Neural Networks after which we describe the quantization scheme we will be using.
In Section \ref{sec:quant-conv-mpc} we provide a self-contained description of our protocol for secure inference, describing the basic building blocks.
We discuss implementation details and present benchmarks in Section \ref{sec:impl-benchm}, and conclude in Section~\ref{sec:conclusions}.

\section{Deep Learning and Quantization}
\label{sec:quantization}

Deep learning models are at the core of many real-world tasks like computer vision, natural language processing and speech recognition.
However, in spite of their high accuracy for many such tasks, their usage on devices like mobile phones, which have tight resource constraints, becomes restricted by the large amount of storage required to store the model and the high amount of energy consumption when carrying out the computations that are typically done over floating-point numbers.
To this end, researchers in the machine learning community have developed techniques that allow weights to be represented by low-width integers instead of the usual 32-bit floating-point numbers, and quantization is recognized to be the most effective such technique when the storage/accuracy ratio is taken into account.

Quantization allows the representation of the weights and activations to be as low as $8$ bits, or even $1$ bit in some cases \cite{DBLP:journals:corr:CourbariauxB16,DBLP:conf:eccv:RastegariORF16}.%
\footnote{Furthermore, some quantization techniques also allow to represent gradients with a small number of bits, which effectively allows for quantized training of neural networks.
  However, this is still in a very early stage, and since we are focused only on inference in this work, we do not present such techniques.}
This is a long-standing research area, with initial works already dating back to the 1990s \cite{fiesler1990weight,balzer1991weight,tang1993multilayer,marchesi1993fast}, and this extensive research body have enabled modern quantized neural networks to have essentially the same accuracy as their full-precision counterparts \cite{courbariaux2015binaryconnect,zhou2016dorefa,gong2014compressing,han2015deep,park2017weighted}.

\subsection{Notation}
\label{sec:notation}
For a value $\b x\in\R^{N_{1}\times N_{2}\times N_{3}}$ we use
$\b x[i,j,c]\in\R$ to denote taking $i$'th value across the first dimension, the
$j$'th value across the second dimension and the $c$'th value across the last
dimension. In a similar way, we might write
$\b x[\cdot, \cdot, c]\in\R^{N_{1}\times N_{2}}$ to denote the matrix obtained
by fixing a specific value for the last dimension. A real value interval is
denoted by $[a,b]$ and a discrete interval by $[a,b]_{\Z}$. We define
\emph{clamping} of a value $x\in\R$ to the interval $[a,b]$, denoted by $\clamp_{a,b}(x)$, by setting
$x\leftarrow a$ if $x < a$, $x\leftarrow b$ if $x>b$ and otherwise
$x\leftarrow x$. (Clamping to a discrete interval is similarly defined.) We
denote by $\N_\ell$ the set $\{1,\ldots,\ell\}$.

\subsection{Deep Learning}
\label{sec:deep-learning-1}
A Convolutional Neural Network (CNN) is an ordered series of non-linear functions $(f_{1},\dots,f_{n})$ where $f_{i}:D_{i-1}\mapsto D_{i}$ is called a \emph{layer}, and where each $D_{i}\in\R^{N_{1}\times\dots\times N_{m_{n}}}$ is a space of tensors.
In practice, and for the networks we consider (ImageNet networks) $D_{0}\in\R^{128\times 128\times 3}$ indicating that inputs are $128$ by $128$ pixel RGB images, and $D_{n}\in\R^{1000}$ indicates that there is 1000 output classes.

We are concerned mainly with the case where $f_{i}$ is a convolution, followed by a Rectified Linear Unit (ReLU).
More precisely, $f_{i}$ can be expressed as $f_{i}(x)=\max(x W + b, 0)$ where $W$ and $b$ are tensors (\emph{weights} and \emph{bias}, respectively), and where $\max$ is applied entrywise.

\emph{Downsampling}, i.e., making the height and width of the output of a layer smaller, can be achieved by applying pooling operations.
Average pooling, for example, goes over windows of some size $w\times h$ in each channel of the input and outputs the average; that is the output $\b y[i,j,c]$ will be the average of a $w\times h$ window centered around $\b x[i,j,c]$, where $\b x$ is the input tensor.

Finally, \emph{batch normalization} \cite{DBLP:conf:icml:IoffeS15} is often employed to speed up training.
The idea is to normalize the inputs to each activation: instead of computing $g(x)$ for input $x$ and activation function $g$, we instead compute $g(y)$ where
\begin{equation}
  \label{eq:bn}
  y = \gamma \left(\frac{x-\mu_{B}}{\sqrt{\sigma^{2}_{B} + \epsilon}}\right) + \beta,
\end{equation}
where $\gamma$, $\beta$ are parameters learned during training, and $\mu_{B}$,
$\sigma_{B}^{2}$ is the mean and variance, respectively, of a batch $B$ of which
$x$ is a member.
Consider an input $y=xW+b$ to $g$.
During inference, we can ``fold'' the batch normalization parameters into the weights, which is done by using $W'$ and $b'$ defined as
\begin{align*}
  W' &= \frac{\gamma W}{\sigma}, & b' = \gamma\left(\frac{b-\mu}{\sigma}\right) + \beta.
\end{align*}
It is straight forward to verify that using $y'=xW'+b'$ yields the expression in Eq.~(\ref{eq:bn}).

\subsection{Quantization of~\cite{DBLP:journals/corr/abs-1712-05877}}
\label{sec:googl-quant-scheme}
The goal of this section is to provide an overview of the quantization technique of Jacob et al.~\cite{DBLP:journals/corr/abs-1712-05877} (see also \cite{krishnamoorthi2018quantizing}) that we will be relying on to get efficient secure inference.
While this particular quantization scheme might not be state of the art, or even the best for all choices of secure inference (e.g., XONN \cite{riazi2019xonn} relies on a different scheme to get efficient inference) we choose this particular scheme for the following reasons:
It is implemented in Tensorflow (more precisely, TFLite \cite{tflite}) and as such we get a user friendly, widely available and well documented way of training models that can be securely evaluated.
The fact that Tensorflow can be used to directly train models for our framework is very handy indeed as it removes the need to develop custom tooling that has little to do with the secure framework itself.
Moreover, Tensorflow provides several pre-trained ImageNet models which provides a very good point of reference for not only our benchmarks, but for future works that wish to compare against us.
Indeed, few if any previous work on secure inference provide pretrained models which makes an accuracy oriented comparison very hard.%
\footnote{This is especially the case if it is not clear exactly how the model was trained and which training and test data was used.}
Our focus here is a particular quantization scheme; for a broader survey, we refer the reader to Guo \cite{guo2018survey}.

We note that this scheme is beneficial for MPC since it simplifies the activations and the arithmetic needed to evaluate a CNN.
However, the original goal of Jacob et al.~was to reduce the size of the models, rather than simplifying the arithmetic or the activations.
Unfortunately, we do not get the benefits in the size reduction since, even if the network can be stored using $8$-bit integers, arithmetic must be done modulo $2^{32}$ and even $2^{64}$ in some cases.

\subsubsection{Quantization and De-Quantization}
The scheme comes in two variants, one for 8-bit integers and another one for 16-bit integers.
In this work we focus in the former, and we provide our description only in that setting.

Let $m\in\R$ and $z\in[0,2^8)_\Z$ and consider the function $\dequant_{m,z}:[0,2^8)_\Z\to\R$ given by $\dequant_{m,z}(x) = m\cdot(x-z)$.
This function transforms the interval $[0,2^8)_\Z$ injectively into the interval $I = [-m\cdot z, m\cdot(2^8-1-z))$ and as such it admits and inverse $\quant_{m,z}$ mapping elements in the image of $\dequant_{m.z}$ into $[0,2^8)_\Z$.
We define the quantization of a number $\alpha\in I$ to be $\quant_{m,z}(\alpha')$, where $\alpha'$ is closest number to $\alpha$ such that $\alpha'$ is in the image of $\dequant_{m,z}$.

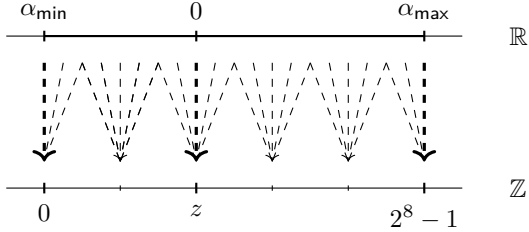
\begin{figure}[h]
  \centering

  \begin{tikzpicture}[node distance=2em]
    \path[draw] (-0.5,0) -- (5.5,0);
    \foreach \x in {0,1,2,3,4,5} {
      \draw (\x,-1pt) -- (\x,1pt);
    }
    \foreach \x in {0,2,5} {
      \draw[thick] (\x,-2pt) -- (\x,2pt);
    }
    \node[below=3pt] at (0,0) {$0$};
    \node[below=3pt] at (2,0) {$z$};
    \node[below=3pt] at (5,0) {$2^8-1$};
    \node[right=1cm] at (5,0) {$\Z$};
    
    \begin{scope}[shift={(0,2)}]
      \path[draw] (-0.5,0) -- (5.5,0);
      \path[draw,thick] (0,0) -- (5,0);
      \foreach \x in {0,2,5} {
        \draw[thick] (\x,-2pt) -- (\x,2pt);
      }
      \node[above=3pt] at (0,0) {$\alpha_{\mathsf{min}}$};
      \node[above=3pt] at (2,0) {$0$};
      \node[above=3pt] at (5,0) {$\alpha_{\mathsf{max}}$};
      \node[right=1cm] at (5,0) {$\R$};
    \end{scope}

    \foreach \x in {0,2,5} {
      \draw[->,very thick,dashed] ([yshift = -10pt] \x,2) -- ([yshift = +10pt] \x,0);
    }
    \foreach \x in {1,3,4} {    
    \draw[->,dashed] ([yshift = -10pt] \x,2) -- ([yshift = +10pt] \x,0);
  }
  \foreach \x/\y in {0.25/0,0.5/0,
    0.5/1,0.75/1,1.25/1,1.5/1,
    1.5/2,1.75/2,2.25/2,2.5/2,
    2.5/3,2.75/3,3.25/3,3.5/3,
    3.5/4,3.75/4,4.25/4,4.5/4,
    4.5/5,4.75/5
} {
    \draw[dashed,thin] ([yshift = -10pt] \x,2) -- ([yshift = +10pt] \y,0);
  }

    \foreach \x in {0.5,0.75,1.25,1.5} {
    \draw[dashed] ([yshift = -10pt] \x,2) -- ([yshift = +10pt] 1,0);
  }

  \end{tikzpicture}

  \caption{Visualization of the Quantization Scheme by Jacob et al. \cite{DBLP:journals/corr/abs-1712-05877}.
  The continuous interval on top is mapped to the discrete interval below, and multiple numbers may map to the same integer due to the rounding.}
  \label{fig:quant}
\end{figure}

The constants $m,z$ above are the parameters of the quantization, and are known as the \emph{scale} and the \emph{zero-point}, respectively.
This quantization method will be applied on a per-tensor basis, i.e. each individual tensor $\b \alpha$ has a single pair $m,z$ associated to it.
These parameters are determined at training time by recording the ranges on which the entries of a given tensor lie, and computing $m,z$ such that the interval $[-m\cdot z, m\cdot(2^8-1-z))$ is large enough to hold these values.
See Figure \ref{fig:quant} for a visualization of this quantization method, and see Jacob et al. \cite{DBLP:journals/corr/abs-1712-05877} for details.

\subsubsection{Dot Products}
\label{sec:dot products}
Computing dot products is a core arithmetic operation in any CNN.
In this section we discuss how to do this with the quantization method described above.

Let $\b \alpha = (\alpha_1,\ldots,\alpha_N)$ and $\b \beta = (\beta_1,\ldots,\beta_N)$ be two vectors of numbers with quantization parameters $(m_1,z_1)$ and $(m_2,z_2)$, respectively.
Let $\gamma = \sum_{i=1}^N\alpha_i\cdot\beta_i$, and suppose that $\gamma$ is part of a tensor whose quantization parameters are $(m_3,z_3)$.
Let $c = \quant_{m_3,z_3}(\gamma)$, $a_i = \quant_{m_1,z_1}(\alpha_i)$ and $b_i = \quant_{m_2,z_2}(\beta_i)$.
It turns out we can compute $c$ from all the $a_i,b_i$ by using integer-only arithmetic and fixed-point multiplication, as shown in the following.

Since $\gamma \approx m_3\cdot(c-z_3)$, $\alpha_i\approx m_1(a_i-z_1)$ and $\beta_i \approx m_2(b_i-z_2)$, it holds that
$$m_3\cdot(c-z_3) \approx \gamma = \sum_{i=1}^N\alpha_i\cdot\beta_i\approx \sum_{i=1}^N m_1\cdot(a_i-z_1)\cdot m_2\cdot (b_i-z_3).$$
Hence, we can approximate $c$ as
\begin{equation}
  \label{eq:2}
  c = z_3 + \frac{m_1\cdot m_2}{m_3}\cdot\sum_{i=1}^N (a_i-z_1)\cdot(b_i-z_2)
\end{equation}

The summation $s = \sum_{i=1}^N (a_i-z_1)\cdot(b_i-z_2)$ involves integer-only arithmetic and it is guaranteed to fit in $16+\log N$ bits, since each summand, being the product of two $8$-bit integers, fits in $16$ bits.
However, since $m = (m_1m_2)/m_3$ is a real, the product $m\cdot s$ cannot be done with integer-only arithmetic.
This product is handled in TFLite by essentially transforming $m$ into a fixed-point number and then performing fixed-point multiplication, rounding to the nearest integer.
More precisely, $m$ is first normalized as $m = 2^{-n}m''$ where $m''\in[0.5,1)$,\footnote{Jacob et al. \cite{DBLP:journals/corr/abs-1712-05877} find that in practice $m\in[0,1)$, which is the reason why such normalization is possible.
  We also confirm this observation in our experiments, although it is not hard to extend this to the general case (in fact, TFLite already supports it).}
and then $m''$ is approximated as $m''\approx 2^{-31}m'$, where $m'$ is a $32$-bit integer.
This is highly accurate since $m''\geq 1/2$, so there are at least $30$ bits of relative accuracy.

Thus, given the above, the multiplication $m\cdot s$ is done by computing the integer product $m\cdot s$, which fits in $64$ bits since both $m$ and $s$ use at most $32$ bits (if $N \le 2^{16}$), and then multiplying by $2^{-n-31}$ followed by a rounding-to-nearest operation.
Finally, addition with $z_3$ is done as simple integer addition.

If the quantization parameters for $\gamma$ were computed correctly, it should be the case, by construction, that the result $c$ lies in the correct interval $[0,2^8)_\Z$.
However, due to the different rounding errors that can occur above, this may not be the case.
Thus, the result obtained with the previous steps is clamped into the interval $[0,2^8)_\Z$.

\subsubsection{Addition of bias}
In the context of CNNs the dot products above will come from two-dimensional convolutions.
However, these operations not only involve dot products but also the addition of a single number, the bias.
In order to handle this in a smooth manner with respect to the dot product above, the scale for the bias is set as $m_1m_2/m_3$ and the zero-point it set to $0$.
This allows the quantized bias to be placed inside the summation $s$, involving no further changes to our description above.

\subsubsection{Other layers}
\label{sec:other-layers}
Other layers like ReLU, ReLU6 or max pooling, which involve only comparisons, can be implemented with relative ease directly on the quantized values, assuming these share the same quantization parameters.
This is because if $\alpha = m(a-z)$ and $\beta = m(b-z)$, then $\alpha\leq\beta$ if and only if $a\leq b$, so the comparisons can be performed directly on the quantized values.

In fact, activations like ReLU6 (which is used extensively in the models we
consider in this work) can be entirely fused into the dot product that precedes
it, as shown in Section 2.4 of \cite{DBLP:journals/corr/abs-1712-05877}. Since
ReLU6 is essentially a clamping operation, it is possible, by carefully picking
the quantization parameters, to make the clamping of the product to the interval
$[0,2^{8})_{\Z}$ \emph{also} take care of the ReLU6 operation. In short, if the
zero-point is 0 and the scale is $6/255$, then we are guaranteed that
$m(q-z)\in[0,6]$ for any $q\in\{0,\dots,2^{8}-1\}$.

On the other hand, mathematical functions like sigmoid must be handled differently.
We will not be concerned with this type of functions in this document since it is the case in practice that ReLU and ReLU6 (or similar activation functions) are typically enough.\footnote{See \cite{DBLP:journals/corr/abs-1712-05877} for a discussion on quantization of mathematical functions.}

\section{Quantized CNNs in MPC}
\label{sec:quant-conv-mpc}

In the previous section we discussed how quantization of neural networks works, or, more specifically, we discussed the quantization scheme by Jacob et al. \cite{DBLP:journals/corr/abs-1712-05877}.
Now, we turn to the discussion about how to implement these operations using MPC.
However, before diving into the details of the protocols we use in this work, we describe the setting we consider for the secure evaluation of CNNs.

\subsection{System and Threat Model}
\label{sec:threat-model}

Like most previous work on secure inference using MPC, we consider a setting where both the model owner and client outsource their model, respectively input to a set of servers that perform that actual secure inference, cf.~the illustration in Figure~\ref{fig:client-server}.

\fullversion{\begin{figure}[h]
  \centering

  \begin{tikzpicture}[node distance=2em]
    \node[alice,minimum size=1cm] (alice) at (0,1) {Model owner};
    \node[bob,minimum size=1cm] (bob) at (0,-1) {Data owner};

    \begin{scope}[xshift=4.5cm,thick,every node/.style={fill=black!10}]
      \draw[pattern=dots,pattern color=black] (0,0) circle (1.9cm);
      \node[draw,circle,inner sep = 6pt] (P1) at (canvas polar cs:angle=90,radius=1cm) {$P_1$};
      \node[draw,circle,inner sep = 6pt] (P2) at (canvas polar cs:angle=210,radius=1cm) {$P_2$};
      \node[pattern=north east lines,draw,circle,inner sep = 6pt] (P3) at (canvas polar cs:angle=330,radius=1cm) {$P_3$};
      \draw[<->] (P2) -- (P3);
      \draw[<->] (P1) -- (P2);
      \draw[<->] (P1) -- (P3);
    \end{scope}
    
    \foreach \x in {1,2} {    \draw[->] (alice) to[out=0,in=180] (P\x);}
    \draw[->] (alice) to[out=0,in=135] (P3);

    \foreach \x in {1,2} {    \draw[->] (bob) to[out=0,in=180] (P\x);}
    \draw[->] (bob) to[out=0,in=225] (P3);    
  \end{tikzpicture}

  \caption{Visualization of the Client-Server model we consider in this work. The model and data owner secret-share their data towards two or three servers, depending on the underlying MPC protocol, who then execute the secure computation and return the result to the clients.}
  \label{fig:client-server}
\end{figure}
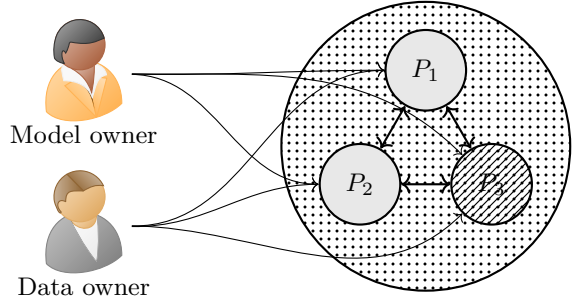
}

We consider a setting of either two or three servers $P_{1}$, $P_{2}$ and $P_{3}$ depending on the setting (honest or dishonest majority) among which one is allowed to be corrupted.
The model and input owner each secret-share their inputs to the servers at the beginning of the protocol execution.
This preserves the privacy of this sensitive information under certain assumptions on the adversarial corruption.
Then, the servers execute a secure multiparty computation protocol to evaluate the quantized model on the given input, obtaining shares of the output, which can then be sent to the party that is supposed to get the classification result.

As we already mentioned previously, our techniques have the crucial feature that virtually \emph{any} secret-sharing-based MPC protocol can be used as the underlying computation engine.
More precisely, let $R$ be either $\Z_{2^k}$ or $\F_p$, we only assume a secret-sharing scheme $\shr{\cdot}$ over $R$ for two or three parties (depending on the setting) withstanding one corruption, allowing local additions $\shr{x+y}\leftarrow\shr{x} + \shr{y}$, together with a protocol for secure multiplication $\shr{x\cdot y}\leftarrow\shr{x}\cdot\shr{y}$.\footnote{Protocols with these features are typically referred to as general-purpose MPC protocols, and any construction that only makes use of these properties is said to be in the arithmetic black-box model.}

\fullversion{
  One can also consider a variant in which the model owner and the data owner are the parties running the actual protocol.
  Fortunately, this can be seen as a particular case of the client-server model with two servers in which the model owner plays one of the servers and the input owner plays the other one.
  This only requires an additional input round to guarantee that the inputs are consistent.

  Finally, another benefit of the client-server model is that it is highly compatible with reactive computation, in which the output of the prediction needs to be used for subsequent operations.
  In this case, since the servers obtain shares of the output, it is straightforward to pipeline this result to another MPC computation.
}

The general-purpose MPC protocols we use in this work can be categorized according to three different dimensions: corruption threshold, type of corruptions and underlying algebraic structure.
For the first dimension we distinguish between two cases: honest vs dishonest majority.
In the former, the adversary is allowed to corrupt strictly less than half of the parties. We instantiate this case with 3 parties and 1 corruption, as that leads to the most efficient protocols.
In the latter case, the adversary can corrupt any number of parties provided at least one party remains honest. We instantiate this setting for 2 parties.
While honest majority protocols impose a stronger security assumption than dishonest majority, they tend to be simpler in their design and thus more efficient.
We further distinguish between passive and active corruptions, where the former means the adversary follows the protocol and the latter allows the adversary to deviate.
Not surprising, actively secure protocols impose an overhead over passively secure ones.
Finally, the algebraic structure on which the computation takes place also plays an important role in terms of efficiency and protocol design, with protocols over $\F_p$ being easier to design and possibly implement, but protocols over $\Z_{2^k}$ providing some efficiency improvements in terms of basic arithmetic and bit-operations \cite{defksv}.

We consider a total of $8$ MPC protocols to support the secure evaluation of the quantized CNNs, corresponding to all the possible combinations of the three dimensions mentioned above (active/passive, honest/dishonest majority and computation modulo a prime or a power-of-two).
Table \ref{fig:protocols} contains an overview of which protocol is used in which security model.
We provide more details on each protocol in Section \ref{sec:mpc-protocols} in the appendix.

\begin{table}[t!]
    \centering
    \setlength{\tabcolsep}{5pt}
    \begin{tabular}{lccc}
      \toprule
      Threshold & & $\Z_{2^k}$ & $\F_p$ \\
      \midrule
      \multirow{2}{*}{$t<n$} & Passive & \otsemitk & \otsemip \\
      & Active & \spdztk & $\lowgear$ \\
      \midrule
      \multirow{2}{*}{$t<n/2$}
      & Passive & \repltk & \replp \\
      & Active & \psrepltk & \psreplp\\
      \bottomrule
    \end{tabular}
  \caption{MPC protocols we use, classified depending on their security level (passive vs. active) and their arithmetic properties (modulo $2^k$ vs. modulo a prime). Names are from MP-SPDZ.}
  \label{fig:protocols}
\end{table}

\subsection{Building Blocks}
\label{sec:building-blocks}

For many applications, the multiplication protocol assumed for $\shr{\cdot}$ is not enough.
In practice, many useful functionalities cannot be nicely expressed in terms of additions and multiplications and therefore, more often than not, researchers end up developing custom protocols for specific applications.
As we argued in Section \ref{sec:related-work}, this also includes the case of secure evaluation of Neural Networks.

In our case, thanks to the quantization scheme by Jacob et al. \cite{DBLP:journals/corr/abs-1712-05877} most of the operations in the evaluation of a quantized Neural Network become additions and multiplications, which are already supported by the MPC protocols we consider here.
Furthermore, the multiplications have a very special structure: they are part of a dot product operation, which can be computed more efficiently for the particular case of passive security with an honest majority.
However, the evaluation still requires non-arithmetic operations like truncations and comparisons, which are more expensive and require specialized subprotocols for their computation.
These, fortunately, can be also implemented in the arithmetic black-box model, that is, making use only of additions and multiplications, which preserves our flexibility when it comes to the underlying MPC protocol.
In what follows we describe the primitives we require in order to integrate the quantized models from Section \ref{sec:quantization} into our secure engine.

\subsubsection{Secure comparison}
\label{sec:secure-comparison}

An important primitive involves comparing two secret-shared values, in order to take certain action depending on which of the two is larger.
However, since revealing which of the two inputs is larger leaks information about the inputs themselves (which is not allowed in many applications), a secure comparison protocol outputs the bit indicating the result of the comparison in secret-shared form.
More precisely, a secure comparison subprotocol allows the parties to compute $\shr{b}\leftarrow \shr{x}\stackrel{?}{<}\shr{y}$, that is, $b=1$ if $x<y$, and $b=0$ otherwise.

Just like the case with truncation, this problem is well motivated and has received enough attention by the community, with many existing proposals providing different trade-offs.
Given this, we may assume the existence of a secure comparison subprotocol, which can be instantiated for example using the constructions from \cite{SCN:CatDeH10} for the field case, and \cite{defksv} for the ring case.
For the special case of replicated secret sharing over $\Ztk$, Mohassel and Rindal \cite{CCS:MohRin18} have proposed a more efficient approach. Comparison is equivalent to extracting the most significant bit of the difference between the two operands. This bit can be computed from the carry bit of adding the three shares, which in turn is possible to achieve by a binary circuit on the local bit decomposition of shares. While this binary circuit has linear complexity in the bit length, it only takes one bit to compute an AND gate in this setting.

\subsubsection{Truncation by a public value}
\label{sec:truncation}

As we have already discussed in the introduction, most existing works in the area of secure inference make use of fixed-point arithmetic, in which a rational number $\alpha$ is approximated by the closest integer to $\alpha\cdot 2^t$, where $t$ is some \emph{fixed} parameter.
To keep the right representation after multiplying two fixed-point numbers, the result must be truncated by $t$ bits, which is a non-linear operation and generates several complexities when done in MPC.

Many solutions have been developed throughout the years for computation over both the field $\F_p$ and the ring $\Z_{2^k}$.
We refer the reader to \cite{SCN:CatDeH10} and \cite{defksv} for details on these.
For the purpose of our work, we assume the existence of a subprotocol that computes $\shr{y}\leftarrow\shr{x}$, where $y = \floor{\frac{x}{2^m}}$, where $m$ is some fixed, \emph{public} parameter.

It is also useful to consider the concept of \emph{probabilistic truncation}.
In this case, instead of obtaining $\shr{y}$ from $\shr{x}$, where $y = \floor{\frac{x}{2^m}}$, a protocol for probabilistic truncation computes $\shr{z}$ where $z = \floor{\frac{x}{2^m}}+u$ and $u$ is some small error.
In practice, $u\in\{0,1\}$, and $u$ is ``biased towards $\round{\frac{x}{2^m}}$'', meaning that $u$ equals $1$ with probability the decimal part of $\frac{x}{2^m}$, which equals $(x\bmod 2^m) / 2^m$.
As an example, if $x=7$ and $m=2$, a protocol for probabilistic truncation would produce either $\floor{\frac{7}{4}} = 1$ or $\floor{\frac{7}{4}}+1 = 2$, where the latter happens with probability $.75$.

Since neural networks tend to be quite resilient to small changes in the activations, which as we will see is the ultimate effect of having probabilistic truncation instead of deterministic (i.e.~exact), this approach should not affect the accuracy of the models substantially, although we do not verify this experimentally.
Furthermore, probabilistic truncation protocols tend to perform much better than deterministic ones, as we show experimentally in Section~\ref{sec:impl-benchm}.
This is because these protocols avoid the usage of expensive binary adders and other similar binary circuits that appear in the deterministic case.
See \cite{SCN:CatDeH10} for some details.
In what follows we introduce some novel protocols for the task of probabilistic truncation over a ring $\Z_{2^k}$.

\paragraph*{Probabilistic truncation over the ring $\Z_{2^k}$.}

Truncation over the ring $\Z_{2^k}$ is considerably more difficult as truncation over $\F_p$, as the latter relies on the fact that division by $2^m$ can be done simply by locally multiplying by the inverse of $2^m$, which is not possible over $\Z_{2^k}$.
The authors in \cite{defksv} and \cite{CCS:MohRin18} propose alternative methods to deal with this issue, but unfortunately their methods require either non-constant number of rounds, or require a large gap between the shares and the secret, which hurts performance.

Instead, we propose a novel method to perform secure truncation over $\Z_{2^k}$, where the shares only need to be one bit larger than the secrets and the number of rounds is constant.
The result may have an error of at most $1$, but this error is biased towards the nearest integer to $x/2^m$, where $x$ is the value being truncated.
In our protocol we assume a method to produce random shared bits, that is, $\shr{b}$ where $b\in\{0,1\}$ is uniformly random and unknown to the adversary.
This can be done in the dishonest majority setting as proposed in \cite{defksv}, or more generally, we can let each party $P_i$ propose a bit $\shr{b_i}$ (which can be checked to be a bit indeed by verifying that $\shr{b_i}\cdot(1-\shr{b_i})$ is $0$) and then the parties XOR these bits together to get one single random bit.

\begin{protocol}{Protocol $\truncpr_{\Ztk}(\shr{x}, m)$}
  \begin{description}
  \item[Pre] $x$ with $\mathsf{MSB}(x)=0$.
  \item[Post] $\shr{x / 2^m}$ rounded according to text
  \end{description}
  Proceed as follows:
  \begin{enumerate}
  \item Generate $k$ random shared bits $\shr{r_i}$ and compute $\shr{r} \leftarrow \sum_i \shr{r_i} \cdot 2^i$.
  \item Open $c \leftarrow \shr{x} + \shr{r}$ and compute $c' \leftarrow (c / 2^m) \bmod 2^{k-m-1}$.
  \item Compute $\shr{b} \leftarrow \shr{r_{k-1}} \oplus (c / 2^{k-1})$.
  \item Output $c - \sum_{i=m}^{k-2} \shr{r_i} \cdot 2^{i-m} + \shr{b} \cdot 2^{k - m - 1}$.
  \end{enumerate}
\end{protocol}

\paragraph*{An improvement for the ring case with three parties, honest majority and passive security.}

By further restricting the setting, more optimizations can be done.
We consider replicated secret sharing over $\Z_{2^k}$ with three parties and passive security.
Our truncation protocol emulates the black-box probabilistic truncation in the
setting of semi-honest computation over a power of two with an honest
majority. Informally, it changes from a symmetric three-party protocol
to a two-party protocol where the third party generates correlated
randomness used by the the other parties. This allows to generate
random values of any bit length at once without the need to generate
such random values bit-wise. The latter is the main cost in black-box
probabilistic truncation because the communication is independent of
the number of bits otherwise.

\begin{protocol}{Protocol $\mathsf{TruncPrSp}_{\Ztk}(\shr{x}, m)$}
  $P_3$ proceeds as follows:
  \begin{enumerate}
  \item Sample random bits $\{r_i\}$ for $i \in [0, k -1]$.
  \item Generate 2-out-of-2 sharings of $r = \sum_i r_i \cdot 2^i$,
    $r_{k-1}$, and $\sum_{i=m}^{k-2} r_i \cdot 2^{i-m}$, and send one
    share to $P_1$ and $P_2$ each.
  \item Generate random $y_1, y_3 \in \Ztk$ and send $y_1$ to $P_1$
    and $y_3$ to $P_2$.
  \item Output $(y_3, y_1)$.
  \end{enumerate}
  $P_1$ and $P_2$ proceed as follows:
  \begin{enumerate}
  \item Convert $\shr{x}$ to a 2-out-of-2 sharing by $P_1$ computing
    $x_1 + x_2$ and $P_2$ proceeding with $x_3$.
  \item Execute $\truncpr_{\Ztk}$ as two-party computation using the
    random values received from $P_3$.
  \item $P_i$: Let $y'_i$ denote the share output by $\truncpr_{\Ztk}$ and
    $\hat{y}_i$ the share received from $P_3$ ($y_1$ or $y_3$). Send
    $y'_i - \hat{y}_i$ to $P_{2-i}$. Denote the received value by
    $\tilde{y}_i$.
  \item $P_1$ outputs $(y_1, y'_1 - \hat{y}_1 + \tilde{y}_1)$, and
    $P_2$ outputs $(y'_2 - \hat{y}_2 + \tilde{y}_2, y_3)$.
  \end{enumerate}
\end{protocol}

For correctness, we have to establish that the parties output a
correct replicated secret sharing of the result. To establish the
correct replicated secret sharing, consider
\begin{align*}
  y'_1 - \hat{y}_1 + \tilde{y}_1 &= y'_1 - \hat{y}_1 + y'_2 - \hat{y}_2 \\
                                 &= \tilde{y}_2 + y'_2 - \hat{y}_2.
\end{align*}
Furthermore,
\begin{align*}
  y_1 + y_3 + y'_1 - \hat{y}_1 + \tilde{y}_1
  &= y_1 + y_3 + y'_1 - \hat{y}_1 + y'_2 - \hat{y}_2 \\
  &= y_1 + y_3 + y'_1 - y_1 + y'_2 - y_2 \\
  &= y'_1 + y'_2,
\end{align*}
which equals the result of $\truncpr_{\Ztk}$ by definition.

Since we only aim for semi-honest security with honest majority, we
have to show that each party does not learn any information about $x$
if all parties follow the protocol. This is trivial for $P_3$ because
they do not receive anything. For $P_1$ and $P_2$, the randomness
received from $P_3$ is independent of $x$. Furthermore, the security
of the two-party $\truncpr_{\Ztk}$ execution follows by the black-box
definition of it. Finally, $\tilde{y}_i$ does not reveal information
because $\hat{y}_{2-i}$ is uniformly random and unknown to $P_i$.

\subsubsection{Truncation by a Secret Value}
\label{sec:trunc-secr-shift}

The truncation protocols we have considered so far assume that the amount of bits to be truncated, $m$, is public.
This is a natural setting and appears for instance in fixed-point multiplication, where $m$ is equal to the amount of bits assigned for the decimal part.
However, as we already argued in Section \ref{sec:quantization}, the quantization scheme we use here differs from traditional fixed-point arithmetic in that the parameters for the discretization are adaptively chosen for each particular layer of the network.
As a side effect, these parameters become information of the model, and therefore they must not be revealed in the computation.
As a result, truncation by secret amounts become necessary.

In this section we present our protocol for truncation by a secret amount.
It takes as input a secret $\shr{x}$ and a shift $m$ represented by $\shr{2^{M-m}}$ where $M$ is some public upper bound on $m$,\footnote{We may alternatively assume that $m$ itself is shared.
  The conversion $\shr{m}\to\shr{2^{M-m}}$ can be achieved then by first bit-decomposing $M-m$ as $\sum_{i}2^i\cdot b_i$, computing shares of each $b_i$ and then outputting $\shr{2^{M-m}} = \prod_i (1 + \shr{b_i} \cdot (2^{2^i} - 1))$.
  However, since in our setting $m$ is known by the client who has the model, it is simpler to assume that the client distributes $\shr{2^{M-m}}$ to begin with.}
and outputs $\shr{y}$ where $y = \floor{\frac{x}{2^m}}$.

\begin{protocol}{Protocol $\truncp_R(\shr{x},\shr{m})$}
  The parties proceed as follows.
  \begin{enumerate}
  \item 
    Compute $\shr{2^{M-m}\cdot x} = \shr{2^{M-m}}\cdot\shr{x}$.
  \item
    Return $\shr{y}\leftarrow\trunc_R(\shr{2^{M-m}\cdot x},M)$.
  \end{enumerate}
\end{protocol}

\paragraph*{Security.}

We informally argue that the shift $m$ used in the protocol remains hidden.
To this end, simply notice that $m$ is provided as input to the MPC protocol in secret-shared form $\shr{m}$.
Then, the multiplication $\shr{2^{M-m}\cdot x} = \shr{2^{M-m}}\cdot\shr{x}$ does not leak anything since we assume the underlying multiplication protocol is secure.
Finally, since we assume the protocol for \emph{public} truncation $\trunc_R$ is secure, the call $\shr{y}\leftarrow\trunc_R(\shr{2^{M-m}\cdot x},M)$ produces correct shares without leaking anything, which implies that $y = \floor{\frac{2^{M-m}x}{2^M}} = \floor{\frac{x}{2^{m}}}$.
The only requirement for this protocol to work is that $(M-m) + \log_2(|x|)$ must be smaller than bit-length of the modulus of the secret sharing scheme since, in this case, it can be seen that $2^{M-m}\cdot x$ does not overflow.

\subsection{Putting it all Together}
\label{sec:putting-it-all}

Using the building blocks that we just described, together with the quantization scheme from Section \ref{sec:googl-quant-scheme}, we can securely evaluate quantized neural networks in an easy way.
As we discussed in Section \ref{sec:quantization}, evaluating a quantized CNN consists mostly of computing the expression in Eq.~(\ref{eq:2}), followed by a clamping procedure.
We describe these computations in this section, along with the other necessary pieces for the evaluation of a quantized CNN.

Recall from Section \ref{sec:quantization} that each weight tensor $\b a$ in a quantized CNN has a scale $m\in\R$ and a zero-point $z\in\Z_{2^8}$ associated to it, such that $\alpha\approx m\cdot(a-z)$ is the actual floating-point numbers corresponding to each $8$-bit integer $a$ in the tensor.
Also, biases are quantized in a similar manner but with a $32$-bit integer instead, a zero point equal to $0$, and a scale that depends on the inputs and output to the layer it belongs to, as explained in Section \ref{sec:dot products}.
We assume that the model owner, who knows all this information, distributes shares to the servers using the scheme described above of the quantized weights and biases of each layer in the network.\footnote{Notice that these values are only $8$-bit long in the clear, but the shares are $64$-bit long.
  The reason is that, although the values are small, the computation must be carried without overflow.
  Therefore we cannot use a modulus that is smaller than the maximum possible intermediate value.}
Also, the zero points associated to each tensor are shared towards the parties.

The scales of the model, on the other hand, are handled in a slightly different way.
Each dot product in the quantized network requires a fixed-point multiplication by a factor $m = (m_1\cdot m_2)/m_3$, borrowing the notation from Section \ref{sec:dot products}.
Recall that this product was handled by writing $m = 2^{-n-31}\cdot m'$, where $m'$ is a $32$-bit integer.

Now, to compute securely the expression in Eq.~(\ref{eq:2}), recall that the parties have shares of the zero points $z_1,z_2,z_3$, the quantized inputs $a_i,b_i$ for $i=1,\ldots,N$, the integer scale $m'$ and the power $2^{L-\ell}$, where $\ell=n+31$ with $2^{-n-31}\cdot m'\approx m = (m_1\cdot m_2)/m_3$, and $L$ is an upper bound on $\ell$.\footnote{Since $n\leq 32$ it suffices to take $M = 63$. In this case, given that $m\geq 31$, it follows that $2^{M-m}\leq 2^{32}$.
  According to Section \ref{sec:trunc-secr-shift}, this imposes the restriction that the modulus for the computation must be at least $32 + 64 = 96$.
  In practice $n$ is smaller than $32$ and this bound can be improved.}
To compute Eq.~(\ref{eq:2}), the parties begin by computing the dot product $\shr{s} = \sum_{i=1}^N(\shr{a_i}-\shr{z_1})\cdot(\shr{b_i}-\shr{z_2})$.
Then, an additional secure multiplication is used in order to compute $\shr{m\cdot s} = \shr{m}\cdot\shr{s}$.
Next, shares of $\round{2^{-n-31}\cdot(m\cdot s)}$ are computed from $\shr{2^{L-\ell}}$ and $\shr{m\cdot s}$ using Protocol $\truncp$ from Section \ref{sec:trunc-secr-shift}, together with the observation that $\big\lfloor{2^{-m} \cdot x}\big\rceil = \big\lfloor{2^{-m} \cdot x + 0.5}\big\rfloor = \big\lfloor{2^{-m} \cdot (x + 2^{m-1})}\big\rfloor$ for breaking a tie by rounding up.

Finally, addition with $\shr{z_3}$ is local, and it is followed by the clamping the result $\shr{x}$ to the interval $[0,2^8)$.
This is done by comparing $\shr{x}$ to the limits (0 and 255) using a secure comparison protocol (see Section \ref{sec:secure-comparison}), followed by an oblivious selection:
If $s \in \{0,1\}$, it holds trivially that $a_s = s \cdot (a_1 - a_0) + a_0$ for arbitrary $a_0, a_1$.

\paragraph*{Other layers.}
Average pooling involves computing $\shr{y}$ from $\shr{x_1},\ldots,\shr{x_n}$, where $y = \round{\frac{1}{n}\cdot\sum_{i=1}^nx_i}$.
This can be achieved using Goldschmidt's algorithm \cite{Goldschmidt},
a widely used iterative algorithm for division. For its usage in the
context of secure multiparty computation, see for example Catrina and Saxena~\cite {FC:CatSax10}. It uses basic arithmetic as well as truncation, both of which we have already discussed.

On the other hand, max pooling requires implementing the max function
securely, which can be easily done by making use of a secure comparison
protocol \cite{SCN:CatDeH10}.

Finally, once shares of the output vector are obtained (raw output, before applying Softmax), several options can be considered.
The parties could open the vector itself towards the input owner and/or data owner so that they compute the Softmax function and therefore learn the probabilities for each label.
However, this would reveal all the prediction vector, which could be undesirable in some scenarios.
Thus, we propose instead to securely compute the argmax of the output array, and return this index, which returns the most likely label since exponentiation is a monotone increasing function.
Previous work, such as SecureML \cite{SP:MohZha17}, replace the exponentiation in the Softmax function with ReLU operations, i.e.~by computing $\relu(x)$ instead of $e^{x}$.
More MPC friendly solutions exist, such as the spherical Softmax~\cite{de2015exploration}, which replaces $e^{x}$ with $x^{2}$.

\section{Implementation and Benchmarking}
\label{sec:impl-benchm}

\begin{table*}[ht!]
  \centering\small
  \setlength{\tabcolsep}{1.5pt}
  \begin{tabular}{ccccccc}
    \toprule
    \multirow{2}{*}{\# sum-of-products} & \multicolumn{2}{c}{mod $2^{k}$} & \multicolumn{2}{c}{mod $p$} & \multicolumn{2}{c}{mod $p$ (active security)} \\
    \cmidrule(lr){2-3} \cmidrule(r){4-5} \cmidrule(lr){6-7} 
& Runtime (s) & Comm. (gb) & Runtime (s) & Comm. (gb) & Runtime (s) & Comm. (gb)  \\\midrule
50\,000 & 0.25 & 0.15 & 1.6 & 0.54 & 8.8 & 4.3 \\
100\,000 & 0.41 & 0.31 & 2.5 & 1.07 & 15.6 & 8.5 \\
150\,000 & 0.57 & 0.46 & 3.6 & 1.59 & 22.5 & 12.8 \\
200\,000 & 0.73 & 0.62 & 4.5 & 2.12 & 29.2 & 17.0 \\
    \midrule
    \textbf{\# of terms} &&&&&& \\
    \midrule               
256 & 0.27 & 0.31 & 1.9 & 1.1 & 9.4 & 5.7 \\
512 & 0.30 & 0.31 & 2.0 & 1.1 & 13.9 & 8.5 \\
768 & 0.33 & 0.31 & 2.3 & 1.1 & 18.5 & 11.4 \\
1024 & 0.36 & 0.31 & 2.4 & 1.1 & 22.9 & 14.3 \\
    \bottomrule
  \end{tabular}
  \caption{Top: running a variable number of sum-of-products of constant length $\ell=512$.
    Bottom: Running $n=100.000$ sum-of-products with variable length.}
  \label{tab:sops-length}
\end{table*}

\pgfplotstablesort[sort key=acc, sort cmp=fixed <]{\ZPassiveTopFive}{data/z-passive-top5}
\pgfplotstablesort[sort key=acc, sort cmp=fixed <]{\ZActiveTopFive}{data/z-active-top5}
\pgfplotstablesort[sort key=acc, sort cmp=fixed <]{\PPassiveTopFive}{data/p-passive-top5}
\pgfplotstablesort[sort key=acc, sort cmp=fixed <]{\PActiveTopFive}{data/p-active-top5}

\begin{figure}[t]
  \centering
  \begin{tikzpicture}
    \begin{axis}[
      scale only axis,
      ylabel = Time (s),
      xlabel = Accuracy (top-5),
      legend pos = north west,
      legend style = {cells={anchor=west}},
      legend entries = {
        {passive, $\Z_{2^{k}}$},
        {passive, $\F_{p}$},
        {active, $\Z_{2^{k}}$},
        {active, $\F_{p}$},
      },
      ]
      \addplot table[x=acc, y=time]
      {\ZPassiveTopFive};
      \addplot table[x=acc, y=time]
      {\PPassiveTopFive};
      \addplot table[x=acc, y=time]
      {\ZActiveTopFive};
      \addplot table[x=acc, y=time]
      {\PActiveTopFive};
    \end{axis}
  \end{tikzpicture}
  \caption{Evaluation times for honest majority protocols. x-axis generally correspond to evaluating larger models.}
  \label{fig:graphs-hm}
\end{figure}
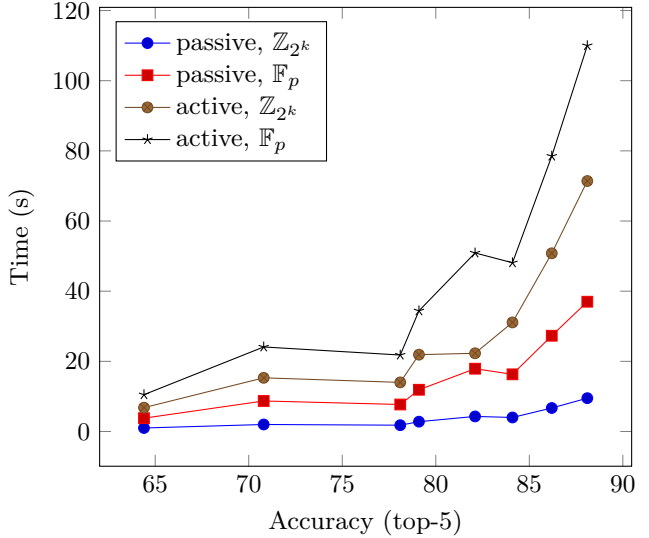

\pgfplotstablesort[sort key=acc, sort cmp=fixed <]{\ZPassiveTopFive}{data/z-dm-passive-top5}
\pgfplotstablesort[sort key=acc, sort cmp=fixed <]{\ZActiveTopFive}{data/z-dm-active-top5}
\pgfplotstablesort[sort key=acc, sort cmp=fixed <]{\PPassiveTopFive}{data/p-dm-passive-top5}
\pgfplotstablesort[sort key=acc, sort cmp=fixed <]{\PActiveTopFive}{data/p-dm-active-top5}

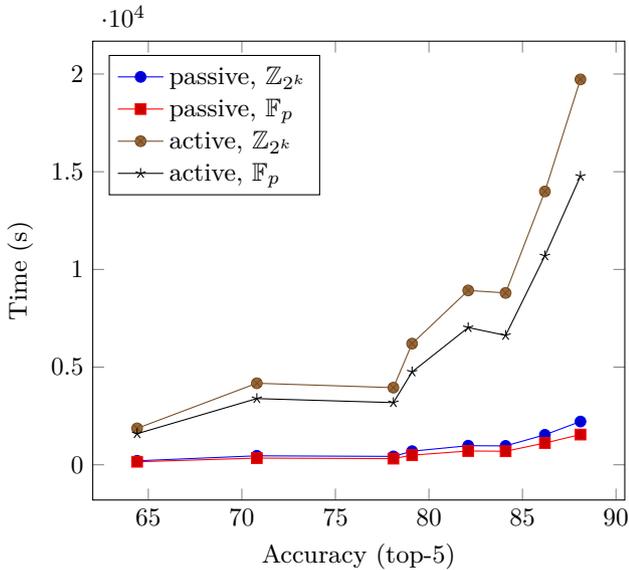
\begin{figure}[t]
  \centering
  \begin{tikzpicture}
    \begin{axis}[
      scale only axis,
      ylabel = Time (s),
      xlabel = Accuracy (top-5),
      legend pos = north west,
      legend style = {cells={anchor=west}},
      legend entries = {
        {passive, $\Z_{2^{k}}$},
        {passive, $\F_{p}$},
        {active, $\Z_{2^{k}}$},
        {active, $\F_{p}$},
      },
      ]
      \addplot table[x=acc, y=time]
      {\ZPassiveTopFive};
      \addplot table[x=acc, y=time]
      {\PPassiveTopFive};
      \addplot table[x=acc, y=time]
      {\ZActiveTopFive};
      \addplot table[x=acc, y=time]
      {\PActiveTopFive};
    \end{axis}
  \end{tikzpicture}
  \caption{Evaluation times for dishonest majority protocols. x-axis generally correspond to evaluating larger models.}
  \label{fig:graphs-dm}
\end{figure}

\begin{table*}[h!]
  \centering
  \setlength{\tabcolsep}{8.5pt}
  \begin{tabular}{ccccccccccccc}
    \toprule
    \multirow{4}{*}{Variant} & \multicolumn{2}{c}{\multirow{3}{*}{Accuracy}} & \multirow{4}{*}{Trunc.} & \multicolumn{4}{c}{Passive Security} & \multicolumn{4}{c}{Active Security} \\
    \cmidrule(lr){5-8} \cmidrule(lr){9-12}
                             & & & & \multicolumn{2}{c}{Dishonest Maj.} & \multicolumn{2}{c}{Honest Maj.} & \multicolumn{2}{c}{Dishonest Maj.}  & \multicolumn{2}{c}{Honest Maj.} \\
    \cmidrule(lr){2-3} \cmidrule(lr){5-6} \cmidrule(lr){7-8} \cmidrule(lr){9-10} \cmidrule(lr){11-12}
                             & Top-1 & Top-5 & & $\Ztk$ & $\F_p$ & $\Ztk$ & $\F_p$ & $\Ztk$ & $\F_p$ & $\Ztk$ & $\F_p$ \\\midrule
    \multirow{2}{*}{V1 0.25\_128} & \multirow{2}{*}{39.5\%} & \multirow{2}{*}{64.4\%} & Prob. & 139.5 & 129.1 & 0.2 & 3.3 & 1264.9 & 1377.8 & 5.3 & 9.0 \\
&&& Exact & 203.2 & 155.0 & 1.0 & 3.8 & 1864.9 & 1592.0 & 6.8 & 10.5 \\
\hline\multirow{2}{*}{V1 0.25\_160} & \multirow{2}{*}{42.8\%} & \multirow{2}{*}{68.1\%} & Prob. & 214.5 & 201.1 & 0.3 & 5.2 & 1997.4 & 2070.4 & 8.1 & 14.3 \\
&&& Exact & 317.7 & 241.2 & 1.4 & 6.1 & 2916.1 & 2432.8 & 10.6 & 16.9 \\
\hline\multirow{2}{*}{V1 0.25\_192} & \multirow{2}{*}{45.7\%} & \multirow{2}{*}{70.8\%} & Prob. & 305.1 & 288.5 & 0.4 & 7.3 & 2827.8 & 2875.0 & 11.8 & 20.3 \\
&&& Exact & 460.6 & 343.1 & 2.0 & 8.7 & 4173.9 & 3389.8 & 15.3 & 24.1 \\
\hline\multirow{2}{*}{V1 0.25\_224} & \multirow{2}{*}{48.2\%} & \multirow{2}{*}{72.8\%} & Prob. & 417.6 & 383.3 & 0.5 & 10.0 & 3825.6 & 3855.3 & 16.1 & 27.3 \\
&&& Exact & 614.1 & 460.8 & 2.9 & 11.8 & 5629.6 & 4574.0 & 20.6 & 32.5 \\
\hline\multirow{2}{*}{V1 0.5\_128} & \multirow{2}{*}{54.9\%} & \multirow{2}{*}{78.1\%} & Prob. & 305.4 & 267.0 & 0.4 & 6.5 & 2731.5 & 2760.4 & 10.9 & 18.5 \\
&&& Exact & 430.1 & 316.3 & 1.8 & 7.7 & 3950.3 & 3183.4 & 14.0 & 21.8 \\
\hline\multirow{2}{*}{V1 0.5\_160} & \multirow{2}{*}{57.2\%} & \multirow{2}{*}{80.5\%} & Prob. & 472.6 & 418.3 & 0.6 & 10.4 & 4331.5 & 4277.2 & 17.7 & 29.8 \\
&&& Exact & 672.1 & 496.5 & 2.9 & 12.5 & 6177.5 & 5006.9 & 22.3 & 34.7 \\
\hline\multirow{2}{*}{V1 0.5\_192} & \multirow{2}{*}{59.9\%} & \multirow{2}{*}{82.1\%} & Prob. & 676.1 & 593.1 & 0.9 & 15.2 & 6194.6 & 6026.6 & 25.5 & 42.7 \\
&&& Exact & 978.0 & 706.3 & 4.3 & 17.9 & 8924.5 & 7025.4 & 32.6 & 50.9 \\
\hline\multirow{2}{*}{V1 0.5\_224} & \multirow{2}{*}{61.2\%} & \multirow{2}{*}{83.2\%} & Prob. & 915.6 & 802.2 & 1.1 & 20.5 & 8446.5 & 8112.9 & 33.7 & 57.7 \\
&&& Exact & 1320.6 & 955.6 & 5.8 & 24.4 & 11962.2 & 9143.3 & 43.7 & 68.2 \\
\hline\multirow{2}{*}{V1 0.75\_128} & \multirow{2}{*}{55.9\%} & \multirow{2}{*}{79.1\%} & Prob. & 485.4 & 421.2 & 0.6 & 9.9 & 4440.8 & 4152.6 & 17.4 & 29.3 \\
&&& Exact & 697.3 & 494.5 & 2.8 & 11.9 & 6203.2 & 4754.5 & 21.9 & 34.4 \\
\hline\multirow{2}{*}{V1 0.75\_160} & \multirow{2}{*}{62.4\%} & \multirow{2}{*}{83.7\%} & Prob. & 775.7 & 662.2 & 1.1 & 15.9 & 7018.5 & 6502.8 & 28.3 & 46.9 \\
&&& Exact & 1075.8 & 779.5 & 4.6 & 19.0 & 9780.5 & 7491.2 & 36.1 & 55.1 \\
\hline\multirow{2}{*}{V1 0.75\_192} & \multirow{2}{*}{66.1\%} & \multirow{2}{*}{86.2\%} & Prob. & 1101.1 & 943.0 & 1.6 & 23.3 & 10053.3 & 9145.2 & 40.5 & 68.0 \\
&&& Exact & 1536.9 & 1114.8 & 6.7 & 27.3 & 13991.2 & 10696.1 & 50.8 & 78.5 \\
\hline\multirow{2}{*}{V1 0.75\_224} & \multirow{2}{*}{66.9\%} & \multirow{2}{*}{86.9\%} & Prob. & 1487.2 & 1276.8 & 2.2 & 31.4 & 13634.5 & 12367.3 & 54.4 & 91.9 \\
&&& Exact & 2135.5 & 1505.8 & 9.6 & 37.4 & 18962.2 & 14370.4 & 69.3 & 107.7 \\
\hline\multirow{2}{*}{V1 1.0\_128} & \multirow{2}{*}{63.3\%} & \multirow{2}{*}{84.1\%} & Prob. & 709.4 & 587.1 & 1.0 & 13.5 & 6381.8 & 5733.0 & 24.9 & 40.9 \\
&&& Exact & 968.5 & 694.1 & 4.0 & 16.3 & 8797.0 & 6624.6 & 31.1 & 48.1 \\
\hline\multirow{2}{*}{V1 1.0\_160} & \multirow{2}{*}{66.9\%} & \multirow{2}{*}{86.7\%} & Prob. & 1101.8 & 928.4 & 1.8 & 21.7 & 10142.0 & 9006.3 & 39.9 & 65.3 \\
&&& Exact & 1528.0 & 1084.0 & 6.5 & 25.9 & 13780.4 & 10357.2 & 49.6 & 76.8 \\
\hline\multirow{2}{*}{V1 1.0\_192} & \multirow{2}{*}{69.1\%} & \multirow{2}{*}{88.1\%} & Prob. & 1581.6 & 1323.9 & 2.6 & 31.5 & 14471.8 & 12778.3 & 57.0 & 95.3 \\
&&& Exact & 2214.8 & 1549.0 & 9.5 & 37.0 & 19725.0 & 14770.0 & 71.4 & 110.0 \\
\hline\multirow{2}{*}{V1 1.0\_224} & \multirow{2}{*}{70.0\%} & \multirow{2}{*}{89.0\%} & Prob. & 2147.3 & 1792.2 & 3.5 & 42.5 & 19691.6 & 17211.3 & 76.9 & 129.0 \\
&&& Exact & 2943.3 & 2101.4 & 13.1 & 50.4 & 26714.3 & 19910.4 & 96.2 & 151.3 \\

    \bottomrule
  \end{tabular}
  \caption{Running time, in seconds, of securely evaluating some of the networks in the MobileNets family, in a LAN network. The first number in variant is the width multiplier and the second is the resolution multiplier.
    Top-1 accuracy measures when the truth label is predicted correctly by the model whereas Top-5 measures when the truth label is among the first 5 outputs of the model.
    Prob. and Exact refer to probabilistic truncation and nearest rounding, respectively.
  }
  \label{fig:mobilenets-time}
\end{table*}

This section discusses our implementation and our performance results.

\subsection{MobileNets}
\label{sec:mobilenets}
Our benchmarks are all performed by evaluating networks of the \emph{MobileNets} type architecture \cite{DBLP:journals:corr:HowardZCKWWAA17}.
A MobileNets network consists of 28 layers with 1000 output classes and are trained on the ImageNet data set \cite{imagenetweb}.
Layers are alternating (with few exceptions at the start and end) \emph{pointwise} convolutions and \emph{depthwise} convolutions.
A pointwise convolution is a regular convolution with a $1\times 1$ filter, while a depthwise convolution can be viewed as a convolution where no summation across output channels occurs.
The size of the network can be adjusted by two hyper parameters: a \emph{width multiplier} $\alpha$ and a \emph{resolution multiplier} $\rho$.
$\alpha$ scales input and output channels, while $\rho$ scales the dimensions of the input image.
Thus $\alpha$ reduces both the model size (as there will be fewer parameters with a smaller $\alpha$) and number of operations, while $\rho$ only scales the number of operations.
In the following we denote a particular model as ``$\mathtt{V1}$ $\alpha\mathtt{\_{}}S$'' where $S$ is the height and width of the input image (thus dependent on $\rho$).
We evaluate the pretrained models which are available on the Tensorflow repository,\footnote{See \url{https://www.tensorflow.org/lite/guide/hosted_models}} and we use their accuracy values.

\paragraph*{Quantizing arbitrary tensorflow models.}
We choose to evaluate the MobileNet models for two reasons: First, they can be considered realistic in the sense that they are expressive enough to solve a wide variety of image related classification tasks.
Thus, the evaluation times we report in this section will correspond to running evaluations of similarly expressive models in practice.
Second, the models are hosted in pre-trained form online which in principle makes our results reproducible (prior work, while sometimes describing the architecture of the models they evaluate, very rarely describe the training process).

However, our technique is by no means limited to running only MobileNet networks.
A quantized model is obtained either by performing quantization aware training%
\footnote{See \url{https://github.com/tensorflow/tensorflow/tree/r1.13/tensorflow/contrib/quantize}}
or by post-quantizing an already trained model%
\footnote{See \url{https://www.tensorflow.org/lite/performance/post_training_quantization}}.
We stress that both these models are implemented entirely by TensorFlow, so no external conversion is needed.

\subsection{Implementation}
\label{sec:benchmarks}

We implement secure inference in the MP-SPDZ framework, which allows us to get timings for all the protocols described in Section \ref{sec:mpc-protocols}.
These protocols run over either a prime $p$ or a ring $\Z_{2^{k}}$.
The prime is 128 bits while the $k$ we use for the ring is 72 bits.
As described in Section \ref{sec:putting-it-all}, these arise because we need some extra space in order for the truncation by a secret shift to be correct.
We arrive at $k=72$ experimentally by computing the sizes of the shift and dot-products needed in the models we evaluate.

We ran all our benchmarks on colocated c5.9xlarge AWS machines, each of which
has 36 cores, 72gb of memory, a 10gpbs link between them and sub-millisecond latency. Throughout this
section, communication is measured per party and all timings include preprocessing.
{%
  Our code has been published as part of MP-SPDZ
  \cite{mp-spdz}.\footnote{The scripts necessary to convert the
    published models and images can be found at
    \url{https://github.com/anderspkd/SecureQ8}.}
}

\subsection{Microbenchmarks}
The main gain in efficiency is obtained by virtue of dot-products, or sums-of-products, being essentially free in some of the protocols we evaluate.
We illustrate the efficiency of this optimization in Table \ref{tab:sops-length}.

Our micro-benchmarks are focused first and foremost on measuring the cost, in
terms of time and communication, of the core operation of any CNN: the
sum-of-product operation (in the MobileNets models, essentially all computations
are convolutions). The top table in Table \ref{tab:sops-length} shows the result of running a
variable number of dot products each of a fixed length, and the bottom table in Table
\ref{tab:sops-length} shows the result of running a fixed number of dot products
with variable length. We choose numbers that reflect realistic sizes for the
convolutions, for example, the largest convolution in the smallest MobileNetsV1
network contains some 60K dot products.

Unsurprisingly, we see a noticeable slowdown for protocols where the communication cost of dot-products depend on the number of terms.
For example, the active security modulo $p$ protocol has a runtime increase of roughly $\times 2.4$, when the number of terms is quadrupled, whereas the passive modulo $2^{k}$ only sees a $\times 1.3$ increase in cost.

\subsection{Full model evaluation}
We evaluate 16 pre-trained V1 MobileNet models of varying sizes.
Each model is evaluated across four different dimensions:
\begin{enumerate}
\item Corruption threshold: We evaluate with both honest and dishonest majority, where the former uses three parties and the latter two.
\item Corruption model: Passive vs.~active security.
\item Algebraic structure: We consider protocols over rings and protocols over fields, with parameters as outlined above.
\item Probabilistic vs.~exact truncation.
\end{enumerate}

Full end-to-end (i.e., with pre-processing) evaluation times for all models in all settings are shown in Table \ref{fig:mobilenets-time}.

\paragraph*{Discussion.}
For the following discussion, we will mainly rely on the two graphs in Figure \ref{fig:graphs-hm} and Figure \ref{fig:graphs-dm}.
Both graphs use the 8 models from Table \ref{fig:mobilenets-time} with $S\in\{128,192\}$.
Figure  \ref{fig:graphs-hm} are evaluation times for protocols with honest majority while Figure  \ref{fig:graphs-dm} are protocols with dishonest majority.

As a first thing, we observe that corruption threshold is the most influential factor in terms of evaluation times.
Indeed, just comparing the y-axis of Figure \ref{fig:graphs-hm} with the y-axis of \ref{fig:graphs-dm} shows that there is a huge difference.
The overhead with respect to moving from honest majority to dishonest majority is as high as 200 times for certain configurations places (active security for $\Z_{2^{k}}$ for $\mathtt{V1\,\,1.0\_192}$, for example).
This large difference would be attributed to the expensive pre-processing that is needed in the dishonest majority case.

On the other hand, moving between different corruption models is relatively cheap.
In this regard, the overhead is in fact more or less the same regardless of the threshold.
I.e., moving from passive to active security only increases the inference time by a factor of between 3 and 30.

We also observe that the choice of algebraic structure---field vs.~ring---provides a performance boost in some cases.
The ring-based protocols mostly outperform the field-based protocols in the passive case, while the reverse is true for active security. This is because we use homomorphic encryption with fields in this case but oblivious transfer with rings, which has a higher communication requirement.
Otherwise, we attribute the difference to the fact that the ring we use is smaller than the field for security requirements.
In particular, operations over $\Z_{2^{72}}$ can be performed by operating on only 72-bits (in particular, support for 128-bit wide types which exist in e.g., GCC can be used), while operating over $\F_{p}$ for $p\approx 2^{128}$ require multiplication of two 128-bit integers \emph{without} overflow even when using Montgomery representation.
Furthermore, we use the faster comparison proposed by Mohassel and Rindal \cite{CCS:MohRin18} in the honest-majority setting with rings.

Finally, we observe, not surprisingly, that inference can be sped up by relying on a less precise method of truncation.
For example, if we consider the first row (model $\mathtt{V1\,\,0.25\_128}$) in Table \ref{fig:mobilenets-time} we see that probabilistic truncation speeds up inference by between 80\% and 15\%.
However, this increase in efficiency comes at the cost of a (possible) decrease in accuracy.
We do not expect that this boost in efficiency will become more pronounced for deeper models, since the exact truncation protocol only depends on the size of the integers being truncated.

\paragraph*{Scaling.}

For protocols that support more than three parties, Table
\ref{table:many} shows how the simplest network scales with up to five
parties. Note we do not use Shamir of replicated secret sharing here
for honest-majority computation. This explains why there is no
ring-based protocol and the discrepancy between the results here and
in Table \ref{fig:mobilenets-time}. The number of corrupted parties is set to
the maximum in
the respective protocols, that is, 2, 3, 4 for dishonest majority, and
1, 1, 2 for honest majority.

\begin{table*}
  \centering
  \begin{tabular}{lcccccccc}
    \toprule
    && \multicolumn{3}{c}{Passive Security} & \multicolumn{3}{c}{Active Security} \\
    \cmidrule(lr){3-5} \cmidrule(lr){6-8}
    & & \multicolumn{2}{c}{Dishonest Maj.} & {Honest Maj.} & \multicolumn{2}{c}{Dishonest Maj.}  & {Honest Maj.} \\
    \cmidrule(lr){3-4} \cmidrule(lr){5-5} \cmidrule(lr){6-7} \cmidrule(lr){8-8}
    & \# parties & $\Ztk$ & $\F_p$ & $\F_p$ & $\Ztk$ & $\F_p$ & $\F_p$ \\ \midrule
    \multirow{3}{*}{Time (s)}
& 3 & 401.1 & 320.9 & 5.5 & 2456.8 & 2255.0 & 24.8 \\
& 4 & 799.6 & 597.4 & 7.4 & 3632.8 & 3063.7 & 36.0 \\
& 5 & 1332.9 & 959.5 & 15.8 & 4814.3 & 3921.0 & 54.7 \\
\midrule
\multirow{3}{*}{Comm. (GB)}
& 3 & 594.8 & 114.8 & 7.3 & 3513.8 & 515.2 & 31.8 \\
& 4 & 1183.0 & 232.2 & 8.2 & 5266.5 & 766.4 & 35.8 \\
& 5 & 1965.1 & 389.2 & 20.7 & 7018.7 & 1016.7 & 68.6 \\

    \bottomrule
  \end{tabular}
  \caption{Time and communication per party for computing
    V1 0.25\_128 with probabilistic truncation.}
  \label{table:many}
\end{table*}

\subsection{Special Truncation}

In order to evaluate the benefit of our special truncation, we have
benchmarked our implementation with and without it against CrypTFlow
\cite{cryptoeprint:2019:1049} using the SecureNN Networks A--D
\cite{DBLP:journals/popets/WaghGC19} as well as the CIFAR10 SqueezeNet
examples in the CrypTFLow
codebase \cite{ezpc}. Networks
A--D are simple networks
consisting of up to about ten layers using only matrix multiplication,
convolution, ReLU, and max-pooling while the CIFAR10 SqueezeNet
involves more than ten of each convolutions and ReLU.
Table \ref{table:cryptflow} shows
our results using the same network setup as previously described. Special
truncation consistently improves over CrypTFlow whereas the results
without are sometimes considerably worse. The improvement is noticeable
because truncation in CrypTFlow simply consists of local operations
whereas we use a protocol for this. The protocol has the advantage
that it does not pose restrictions on the secret value whereas the
method in CrypTFlow requires that the most significant $s$ bits of the
secret are zero for a statistical security parameter $s$.
Without special truncation, we rely on a protocol that requires $k$
random bits to mask a $k$-bit value. The generation of random bits in
turn requires at least $k$ bit in communication, which makes the
overall communication quadratic in $k$. Special truncation however has
communication cost linear in $k$.
Note also that we
use comparisons as in ABY$^3$ \cite{CCS:MohRin18}, which is
comparable to the approach in CrypTFlow in that the protocol is very
specific to the security model and computation domain.

\begin{table}
  \centering
  \begin{tabular}{llrrrrr}
    \toprule
    && A & B & C & D & SN \\
    \midrule
    \multirow{3}{*}{Time} &
    CrypTFlow & 16 & 57 & 90 & 24 & 622 \\
    & Ours w/o \textsf{TruncPrSp} & 23 & 67 & 122 & 22 & 2099 \\
    & Ours w/ \textsf{TruncPrSp} & 13 & 18 & 49 & 15 & 484 \\
    \midrule
    \multirow{3}{*}{Comm} &
    CrypTFlow & 1.9 & 6.2 & 15.3 & 2.2 & 187 \\
    & Ours w/o \textsf{TruncPrSp} & 2.3 & 28.1 & 44.3 & 3.9 & 512 \\
    & Ours w/ \textsf{TruncPrSp} & 1.1 & 2.6 & 7.0 & 1.1 & 59 \\
    \bottomrule
  \end{tabular}
  \caption{Time (in ms) and total communication (in MB) for SecureNN A--D and CIFAR10
    SqueezeNet networks.}
  \label{table:cryptflow}
\end{table}

We have also implemented CrypTFlow's ImageNet examples with and without
special truncation. The results can be found in Table
\ref{table:imagenet}. Note the results were obtained with the optimal
number of threads for both frameworks, which is 32 for MP-SPDZ and 8
for CrypTFlow.

\begin{table}
  \centering
  \begin{tabular}{llrrr}
    \toprule
    && SN & RN-50 & DN-121 \\
    \midrule
    \multirow{3}{*}{Time} &
    CrypTFlow & 10.9 & 26.9 & 37.2 \\
    & Ours w/o \textsf{TruncPrSp} & 2.5 & 18.9 & 19.8 \\
    & Ours w/ \textsf{TruncPrSp} & 0.6 & 4.7 & 3.7 \\
    \midrule
    \multirow{3}{*}{Comm} &
    CrypTFlow & 2.6 & 6.9 & 10.5 \\
    & Ours w/o \textsf{TruncPrSp} & 7.4 & 53.0 & 60.3 \\
    & Ours w/ \textsf{TruncPrSp} & 0.8 & 3.8 & 4.6 \\
    \bottomrule
  \end{tabular}
  \caption{Time (in s) and total communication (in GB) for SqueezeNet,
    ResNet-50, and DenseNet-121 classification for ImageNet.}
  \label{table:imagenet}
\end{table}

\section{Conclusions}
\label{sec:conclusions}
We show that it is possible to securely evaluate large and realistic networks, so called ImageNet networks, using more-or-less existing MPC protocols.
Moreover, the networks we evaluate are unmodified and can be trained using standard Tensorflow or any other framework which supports the type of quantization discussed (which currently includes both PyTorch and MXNet).
This work thus provides a very appealing approach to secure evaluation from an end-users perspective:
First, because standard MPC suffices, it is possible to choose from a wider array of threat models than previous works allow.
While the passive security honest majority setting is by far the most efficient, our benchmarks still provide an interesting insight into the exact trade-off one wants secure inference against dishonest majority.
Second, the fact that models directly output by Tensorflow can be evaluated \emph{without} modification, means that model designers can remain oblivious to the secure framework.
However, we also saw that choices of more specialized protocols, such as our special probabilistic truncation, can be beneficial if one wants a trade-off in terms of prediction accuracy and speed.

\bibliographystyle{plain}
\bibliography{abbrev0,crypto,references}

\begin{appendix}
  \section{Related work on secure inference}
\label{sec:related-work-secure}

Secure evaluation of Neural Networks can be traced back to at least the work by Orlandi et al.~\cite{barni2006privacy,orlandi2007oblivious} which present a solution based on HE techniques.
Several later works rely on HE techniques either in full or in part.
CryptoNets \cite{DBLP:conf:icml:Gilad-BachrachD16} use Leveled Homomorphic Encryption (LHE), which necessitates bounding the number of operations a priori.
In addition, HE only permits evaluation of polynomials and as such cannot compute e.g., the Rectified Linear activation functions (the function $x\mapsto\max(0,x)$) and the authors therefor rely on the approximation $x\mapsto x^{2}$.
However, and as pointed out by Gilad-Bachrach et al. \cite{DBLP:conf:icml:Gilad-BachrachD16}, such an approximation makes training difficult for larger networks, the issue being that the derivative of $x^{2}$ is unbounded.
Chabanne et al. \cite{EPRINT:CWMMP17} improve upon CryptoNets by evaluating networks with 6 hidden layers (as opposed to only 2 as in Gilad-Bachrach et al.).
More recently, Bourse et al.~\cite{C:BMMP18} obtain faster evaluation albeit for a smaller network (one and two hidden layers) by combining FHE and Discretized Neural Networks (i.e., networks where weights are in $\set{1,-1}$).

One of the downsides of HE based solutions are their inefficiency and inability to handle common activation functions.
Gazelle \cite{USENIX:JuvVaiCha18} combines garbled circuits (GC) with additive HE (AHE) in order to obtain a more efficient system.
The boost in efficiency is attributed to an efficient method of switching between the AHE scheme and a GC, where the former is used to compute convolutions and fully connected layers, while the latter is used to compute the network's activation functions.

The idea of using multiple different protocols to achieve faster predictions have been used before \cite{USENIX:JuvVaiCha18}.
MiniONN \cite{CCS:LJLA17} develops a technique for turning a pretrained model into an oblivious one, which can be evaluated using a mix of HE, additive secret sharing and GC.
Chameleon \cite{ASIACCS:RWTSSK18}, which is an extension of the ABY framework by Demmler et al.~\cite{NDSS:DemSchZoh15}, likewise use secret sharing for matrix operations and GC for activation functions.
More recently, ABY3 by Mohassel and Rindal \cite{CCS:MohRin18}, also benchmark secure evaluation (albeit the authors do not implement full inference) in a framework that relies on a mix of secret sharing, boolean (i.e., GMW) and garbled circuits.

Finally, like solutions relying purely on HE have been considered before, so has solutions that rely purely on GC or MPC; the latter of which is most relevant to this work.
DeepSecure \cite{DBLP:conf:dac:RouhaniRK18} is perhaps the first work to take a pure GC based approach for evaluating Neural Networks.
More recently XONN \cite{riazi2019xonn} builds a very efficient GC based solution by noting that Binarized Neural Networks \cite{hubara2016binarized} (i.e., networks with weights that are bits) can be evaluated very efficiently. XONN shows that evaluating deep networks ($>$ 20 layers) is possible.
A different approach is taken by Ball et al. \cite{cryptoeprint:2019:338} where the authors use the arithmetic garbling technique of Ball et al.~\cite{CCS:BalMalRos16} to evaluate Neural Networks.
Pure MPC based solutions have been studied in SecureML \cite{SP:MohZha17}, which employs a three-party honest majority protocol.
A major performance boost in SecureML can be attributed to the way fixed point arithmetic is handled, where the authors show that it is possible to just have parties perform the truncation locally.
SecureNN \cite{DBLP:journals/popets/WaghGC19} can be seen as an extension of SecureML where both three- and four-party protocols (both with one corrupted party) are used.
Concurrently to this work, CrypTFlow \cite{cryptoeprint:2019:1049} builds a system on top of SecureNN that is capable of evaluating very large networks ($>$100 layers) in reasonable time.
Another very attractive feature of CrypTFlow is that it provides a more complete framework that accepts standard Tensorflow trained models as input (hence the name).

  \section{MPC Protocols}
\label{sec:mpc-protocols}

\subsection{Dishonest Majority}
Protocols in the dishonest majority setting are often harder to develop and they are also more complex than honest majority ones.
They are typically based in additive secret sharing and use authentication tags for active security to ensure that the openings of shared values are done correctly.

\noindent$\bullet$ \spdztk:
This is the first actively secure protocol over $\Z_{2^k}$ in the dishonest majority setting, and it was proposed initially by Cramer et al. \cite{C:CDESX18} and implemented subsequently by Damg{\aa}rd et al.~\cite{defksv}.
This protocol can be seen as an extension of \mascot~\cite{CCS:KelOrsSch16} (itself being an extension of SPDZ, hence the name).
Multiplications in \spdztk\ are handled using multiplication triples, which are preprocessed using oblivious transfer like in \mascot.
Authentication is handled like in SPDZ, but with an addition that allows this method to work over $\Z_{2^k}$ which consists of working over the ring $\Z_{2^{k+s}}$ and using the upper $s$ bits for authentication.

\noindent$\bullet$ \otsemitk, \otsemip:
These protocols denote cut-down versions of $\spdztk$ and \mascot,
respectively. In particular, they omit the usage of
authentication tags and the so-called ``sacrifice'' where two triples
are checked against each other and only of them can subsequently used
in the protocol. There essentially remains the generation of
multiplication triples using OT.

\noindent$\bullet$ $\lowgear$:
This is an actively secure protocol for computation modulo a prime. It
uses semi-homomorphic encryption based on learning with errors. See
Keller et al.~\cite{EC:KelPasRot18} for details.

\subsection{Honest Majority}
Honest majority protocols are typically developed using Shamir Secret Sharing (for an arbitrary number of parties) or Replicated Secret Sharing (for small number of parties).
Since we consider only a small number of servers we focus on the replicated SS instantiations.

\noindent$\bullet$ \repltk, \replp:
This protocol secret-shares a value $x$ among three parties $P_1,P_2,P_3$ by letting each $P_i$ have random pairs $(x_i,x_{i+1})$ (indexes wrap around modulo $3$) subject to $x\equiv x_1+x_2+x_3\bmod M$, where $M=2^k$ for the ring case and $M=p$ for the field case.
The most efficient passively secure multiplication protocol to date is the one presented by Araki et al.~\cite{CCS:AFLNO16}, where the total communication involves $3$ ring elements.

\noindent$\bullet$ \psreplp:
This protocol by Lindell and Nof \cite{CCS:LinNof17} extends \replp{} to active security by preprocessing potentially incorrect triples and proceeding to the online phase using these, optimistically, checking their correctness at the end of the execution using sacrificing techniques.

\noindent$\bullet$ \psrepltk:
This protocol by Eerikson et al.~\cite{cryptoeprint:2019:164} is an extension of the one by Lindell et al. \cite{CCS:LinNof17} to the ring setting.
This is achieved by incorporating ideas by Cramer et al.~\cite{C:CDESX18} in order to adapt the post-sacrifice step by Lindell et al. to the ring $\Z_{2^k}$.

  \section{Extended results}
\label{sec:extended-results}

\paragraph*{Communication and preprocessing}

\begin{table*}[h!]
  \centering
  \setlength{\tabcolsep}{8.5pt}
  \begin{tabular}{cccccccccccc}
    \toprule
    \multirow{4}{*}{Variant} & \multicolumn{2}{c}{\multirow{3}{*}{Accuracy}} & \multirow{4}{*}{Trunc.} & \multicolumn{4}{c}{Passive Security} & \multicolumn{4}{c}{Active Security} \\
    \cmidrule(lr){5-8} \cmidrule(lr){9-12}
                             & & & & \multicolumn{2}{c}{Dishonest Maj.} & \multicolumn{2}{c}{Honest Maj.} & \multicolumn{2}{c}{Dishonest Maj.}  & \multicolumn{2}{c}{Honest Maj.} \\
    \cmidrule(lr){2-3} \cmidrule(lr){5-6} \cmidrule(lr){7-8} \cmidrule(lr){9-10} \cmidrule(lr){11-12}
                             & Top-1 & Top-5 & & $\Ztk$ & $\F_p$ & $\Ztk$ & $\F_p$ & $\Ztk$ & $\F_p$ & $\Ztk$ & $\F_p$ \\\midrule
    \multirow{2}{*}{V1 0.25\_128} & \multirow{2}{*}{39.5\%} & \multirow{2}{*}{64.4\%} & Prob. & 199.2 & 37.1 & 0.1 & 3.4 & 1748.4 & 282.4 & 2.5 & 8.7 \\
&&& Exact & 296.3 & 44.6 & 1.0 & 4.0 & 2578.2 & 335.3 & 4.7 & 10.3 \\
\hline\multirow{2}{*}{V1 0.25\_160} & \multirow{2}{*}{42.8\%} & \multirow{2}{*}{68.1\%} & Prob. & 311.2 & 58.0 & 0.1 & 5.4 & 2731.2 & 423.9 & 3.8 & 13.6 \\
&&& Exact & 462.8 & 69.6 & 1.5 & 6.2 & 4027.1 & 511.9 & 7.3 & 16.1 \\
\hline\multirow{2}{*}{V1 0.25\_192} & \multirow{2}{*}{45.7\%} & \multirow{2}{*}{70.8\%} & Prob. & 447.5 & 83.4 & 0.1 & 7.7 & 3927.0 & 600.0 & 5.4 & 19.6 \\
&&& Exact & 665.6 & 100.2 & 2.2 & 8.9 & 5792.2 & 723.3 & 10.4 & 23.2 \\
\hline\multirow{2}{*}{V1 0.25\_224} & \multirow{2}{*}{48.2\%} & \multirow{2}{*}{72.8\%} & Prob. & 608.5 & 113.3 & 0.2 & 10.5 & 5339.9 & 811.3 & 7.3 & 26.6 \\
&&& Exact & 905.3 & 136.3 & 3.0 & 12.2 & 7877.7 & 987.2 & 14.2 & 31.5 \\
\hline\multirow{2}{*}{V1 0.5\_128} & \multirow{2}{*}{54.9\%} & \multirow{2}{*}{78.1\%} & Prob. & 438.0 & 79.1 & 0.1 & 6.9 & 3834.5 & 581.7 & 5.8 & 18.5 \\
&&& Exact & 631.8 & 94.0 & 1.9 & 7.9 & 5492.6 & 687.4 & 10.2 & 21.7 \\
\hline\multirow{2}{*}{V1 0.5\_160} & \multirow{2}{*}{57.2\%} & \multirow{2}{*}{80.5\%} & Prob. & 684.6 & 123.4 & 0.2 & 10.7 & 5993.7 & 899.7 & 9.0 & 28.8 \\
&&& Exact & 987.4 & 146.8 & 3.0 & 12.4 & 8583.4 & 1075.6 & 15.9 & 33.8 \\
\hline\multirow{2}{*}{V1 0.5\_192} & \multirow{2}{*}{59.9\%} & \multirow{2}{*}{82.1\%} & Prob. & 984.8 & 177.6 & 0.3 & 15.5 & 8621.7 & 1286.9 & 12.9 & 41.5 \\
&&& Exact & 1420.7 & 211.3 & 4.4 & 17.9 & 12349.8 & 1533.4 & 22.9 & 48.7 \\
\hline\multirow{2}{*}{V1 0.5\_224} & \multirow{2}{*}{61.2\%} & \multirow{2}{*}{83.2\%} & Prob. & 1339.4 & 241.6 & 0.3 & 21.0 & 11725.6 & 1744.6 & 17.5 & 56.4 \\
&&& Exact & 1932.6 & 287.4 & 5.9 & 24.3 & 16799.2 & 2079.1 & 31.2 & 66.3 \\
\hline\multirow{2}{*}{V1 0.75\_128} & \multirow{2}{*}{55.9\%} & \multirow{2}{*}{79.1\%} & Prob. & 716.9 & 125.9 & 0.2 & 10.3 & 6264.3 & 916.3 & 10.0 & 29.2 \\
&&& Exact & 1007.6 & 148.3 & 2.9 & 11.9 & 8750.4 & 1074.9 & 16.7 & 34.1 \\
\hline\multirow{2}{*}{V1 0.75\_160} & \multirow{2}{*}{62.4\%} & \multirow{2}{*}{83.7\%} & Prob. & 1120.8 & 196.7 & 0.3 & 16.1 & 9793.1 & 1428.3 & 15.6 & 45.7 \\
&&& Exact & 1574.8 & 231.7 & 4.6 & 18.6 & 13676.4 & 1692.3 & 26.1 & 53.2 \\
\hline\multirow{2}{*}{V1 0.75\_192} & \multirow{2}{*}{66.1\%} & \multirow{2}{*}{86.2\%} & Prob. & 1612.4 & 283.0 & 0.4 & 23.2 & 14089.1 & 2044.4 & 22.4 & 65.8 \\
&&& Exact & 2266.2 & 333.5 & 6.6 & 26.8 & 19680.3 & 2432.0 & 37.5 & 76.6 \\
\hline\multirow{2}{*}{V1 0.75\_224} & \multirow{2}{*}{66.9\%} & \multirow{2}{*}{86.9\%} & Prob. & 2193.2 & 385.0 & 0.5 & 31.5 & 19163.5 & 2783.4 & 30.5 & 89.5 \\
&&& Exact & 3082.9 & 453.7 & 8.9 & 36.5 & 26773.0 & 3294.5 & 51.0 & 104.2 \\
\hline\multirow{2}{*}{V1 1.0\_128} & \multirow{2}{*}{63.3\%} & \multirow{2}{*}{84.1\%} & Prob. & 1035.9 & 177.6 & 0.2 & 13.7 & 9037.9 & 1286.1 & 15.2 & 41.1 \\
&&& Exact & 1423.5 & 207.6 & 3.9 & 15.9 & 12352.1 & 1514.8 & 24.2 & 47.5 \\
\hline\multirow{2}{*}{V1 1.0\_160} & \multirow{2}{*}{66.9\%} & \multirow{2}{*}{86.7\%} & Prob. & 1619.6 & 277.6 & 0.4 & 21.5 & 14128.8 & 2009.9 & 23.7 & 64.2 \\
&&& Exact & 2224.9 & 324.4 & 6.1 & 24.8 & 19306.2 & 2361.8 & 37.7 & 74.3 \\
\hline\multirow{2}{*}{V1 1.0\_192} & \multirow{2}{*}{69.1\%} & \multirow{2}{*}{88.1\%} & Prob. & 2330.3 & 399.4 & 0.5 & 30.9 & 20328.5 & 2889.9 & 34.1 & 92.4 \\
&&& Exact & 3201.9 & 466.8 & 8.7 & 35.7 & 27782.7 & 3400.7 & 54.2 & 106.9 \\
\hline\multirow{2}{*}{V1 1.0\_224} & \multirow{2}{*}{70.0\%} & \multirow{2}{*}{89.0\%} & Prob. & 3169.9 & 543.5 & 0.7 & 42.0 & 27652.5 & 3928.3 & 46.4 & 125.8 \\
&&& Exact & 4356.2 & 635.1 & 11.9 & 48.6 & 37798.3 & 4615.2 & 73.7 & 145.4 \\

    \bottomrule
  \end{tabular}
  \caption{Communication complexity, in Gigabytes, of securely evaluating some of the networks in the MobileNets family, in a LAN network. The first number in variant is the width multiplier and the second is the resolution multiplier.
    Top-1 accuracy measures when the truth label is predicted correctly by the model whereas Top-5 measures when the truth label is among the first 5 outputs of the model.
  }
  \label{fig:mobilenets-comm}
\end{table*}

Table \ref{fig:mobilenets-comm}, analogous to Table \ref{fig:mobilenets-time}, presents the communication, in Gigabyes, used by the protocols we consider when evaluating different ImageNet models.
As noted in Section \ref{sec:impl-benchm}, dishonest majority protocols require a great deal of preprocessing material in order to evaluate a network, which can be seen by the large differences between the values in columns corresponding to dishonest majority, with respect to honest majority.
Interestingly, the protocol over $\F_{p}$ are cheaper with active security than the protocol over $\Z_{2^{k}}$.
This is likely due to the fact that preprocessing in $\F_{p}$ (with active security) is more communication efficient, than the protocol over $\Z_{2^{k}}$, as illustrated in Table \ref{fig:mobilenets-comm}.

\paragraph*{WAN Benchmarks}

We have also run the smallest model in a WAN setting where each party
is located on a different continent. For computation over rings with
probabilistic truncation, the timings range from 110 seconds for
passive honest-majority computation to 28,000 seconds for active
dishonest-majority computation.

\end{appendix}

\end{document}